\documentclass[a4paper,11pt]{article}

\usepackage{bm}
\usepackage{anysize}
\usepackage{amsthm}
\usepackage{amssymb}
\usepackage{amsmath}
\usepackage{fancybox}
\usepackage{bbm}
\usepackage{amsfonts}
\usepackage[T1]{fontenc}
\usepackage{latexsym}
\usepackage{makeidx}
\usepackage{cite}
\usepackage[numbers]{natbib}

\usepackage[pdftex]{graphicx}
\usepackage[pdftex]{color}

\usepackage[all]{xy}

\usepackage{stackengine}
\stackMath

\newtheorem{teo}{Theorem}
\newtheorem{lem}{Lemma}
\newtheorem{pro}{Proposition}

\newtheorem{defi}{Definition} 

\newcommand*{\m}[1]{\underline{#1}}
\newcommand{\un}[1]{\underline{#1}}

\newcommand{\fd}{\rightarrow}

\newcommand{\inc}{\subset}

\newcommand{\iso}{\cong}

\newcommand{\onda}{\widetilde}

\newcommand{\al}{\alpha}

\newcommand{\lan}{\lambda}
\newcommand{\fhi}{\varphi}
\newcommand{\del}{\delta}
\newcommand{\Del}{\Delta}

\newcommand{\Gam}{\Gamma}
\newcommand{\Lam}{\Lambda}
\newcommand{\cit}{\theta}

\newcommand{\Om}{\Omega}

\newcommand{\N}{\mathbb{N}}

\newcommand{\R}{\mathbb{R}}
\newcommand{\C}{\mathbb{C}}

\newcommand{\Sa}{\mathbb{S}}

\newcommand{\pa}{\partial}

\newcommand{\bal}{\boldsymbol \alpha}
\newcommand{\bomega}{\boldsymbol \omega}

\newcommand{\bbe}{\boldsymbol \beta}

\newcommand{\sgn}{\mbox{sgn}}

\newcommand{\St}{\textup{St}}
\newcommand{\el}{\ell}

\newtheorem{remark}{Remark}[section]

\def\pf{\par\noindent {\em Proof.}~\par\noindent}

\def\lim{\mathop{\mbox{\normalfont lim}}\limits}

\def\pf{\par\noindent {\em Proof. }}

\def\pa{\partial}

\def\ve{\varepsilon}

\begin{document}

\date{}

\title{Pizzetti and Cauchy formulae for higher dimensional surfaces:\\ a distributional approach}
\small{
\author
{Al\'i Guzm\'an Ad\'an, Frank Sommen}
\vskip 1truecm
\date{\small  Clifford Research Group, Department of Mathematical Analysis, Faculty of Engineering
and Architecture, Ghent University, Krijgslaan 281, 9000 Gent, Belgium. \\
{\tt Ali.GuzmanAdan@UGent.be}, \hspace{.3cm}{\tt Franciscus.Sommen@UGent.be} }

\maketitle

\begin{abstract} 

{In this paper, we study Pizzetti-type formulas for Stiefel manifolds and Cauchy-type formulas for the tangential Dirac operator from a distributional perspective. First we illustrate a general distributional method for integration over manifolds in $\R^m$ defined by means of $k$ equations $\fhi_1(\m{x})=\ldots=\fhi_k(\m{x})=0$. Next, we apply this method to derive an alternative proof of the Pizzetti formulae for the real Stiefel manifolds  $\textup{SO}(m)/\textup{SO}(m-k)$. Besides, a distributional interpretation to invariant oriented integration is provided. In particular, we obtain a distributional Cauchy theorem for the tangential Dirac operator on an embedded $(m-k)$-dimensional smooth surface. }



\noindent

\vspace{0.3cm}

\small{ }

\noindent
\textbf{}
\end{abstract}


\section{Introduction}
Integration over surfaces of co-dimension $k$ embedded in an $m$-dimensional Euclidean space ($k\leq m$) plays an important role in contemporary mathematics and physics. There are two basic ways of describing smooth $(m-k)$-surfaces in $\R^m$. We either use a parametric definition as the image of a map $\m{x}(\cdot):V'\inc\R^{m-k}\fd \R^m$, $(u_1,\ldots, u_{m-k})\mapsto \m{x}(u_1,\ldots, u_{m-k})$, or we use an implicit definition by means of equations $\fhi_1(\m{x})=\ldots=\fhi_k(\m{x})=0$.
Most integration methods have been traditionally developed and used for a parametric definition. For instance, the non-oriented integral of a function $f$ over an $(m-k)$-surface $\Sigma\inc\R^m$, parametrically defined as before, is given by
\[\int_\Sigma f \, dS = \int_{V'} f(\m{x}(\m{u})) \sqrt{\det G} \,  du_1\ldots du_{m-k},\]
where $G=\left\{\langle \pa_{u_j}[\m{x}], \pa_{u_k}[\m{x}]\rangle\right\}_{j,k=1\ldots,m-k}$ is the so-called Gram matrix of the coordinate tangent vectors. In a similar way, oriented integration over parametric surfaces can be computed by means of differential forms.

On the other hand, some methods of calculation are necessary also in the case where an $(m-k)$-surface is given by means of equations. 
This is the case for integrals over manifolds with a group invariant structure (homogenous spaces) such as $\textup{O}(m)$, $\textup{SO}(m)$, $\textup{U}(m)$, $\textup{SU}(m)$, $\textup{SO}(m)/\textup{SO}(m-k)$,$\textup{U}(m)/\textup{U}(m-k)$, etc. They appear in harmonic analysis \cite{MR0084104, MR3395222, MR3591150}, representation theory \cite{MR0425519}, quantum field theory and random matrix theory \cite{MR1196901, MR1299521}, and many other fields. Integration over this kind of manifolds is normally taken with respect to the Haar measure, i.e. the unique invariant  measure on a compact Lie group. In a similar way, one uses the induced Haar measure on cosets. { In the case of implicitely defined manifolds, the Haar measure can be  written in terms of the Dirac distribution $\del$.}

For example, the group $\textup{O}(m)$ is the  $\frac{m(m-1)}{2}$-dimensional manifold, embedded in the $m^2$-dimensional Euclidean space, that is defined by the $\frac{m(m+1)}{2}$ equations 
\begin{align*}
\langle\m{v}_j,\m{v}_k\rangle&=\del_{jk}, & j,k&=1,\ldots,m,
\end{align*}
where $\m{v}_1, \ldots, \m{v}_m\in\R^m$, $\langle\cdot,\cdot\rangle$ denotes the Euclidean inner product in $\R^m$ and $\del_{jk}$ is the Kronecker symbol.  { The Haar measure on O$(m)$ can be written as the delta function $\prod_{1\leq j\leq k\leq m} \del(\langle\m{v}_j,\m{v}_k\rangle -\del_{jk})$ times the Lebesgue measure in $\R^{m^2}$. In a similar way, the Haar measure on U$(m)$ can be described in terms of the Dirac delta distribution, see e.g. \cite{MR1863143, MR3556414}.}


{The above distributional technique is applied not only to algebraic Lie groups (and cosets) but to more general manifolds.
For instance, in co-dimension $k=1$, we find the following formula by L. H\"ormander, see \cite[Theorem 6.1.5]{MR1996773} (although an earlier antecedent in physics can be found in \cite{MR0116523}),}
\begin{equation}\label{HorFor}
\int_{\Sigma} \, f\,dS=\int_{\R^m} \del(\fhi(\underline{x}))\, \|\pa_{\underline{x}}[\fhi](\underline{x})\|\, f(\m{x})\; dV_{\underline{x}}.
\end{equation}
Here  $dV_{\underline{x}}=dx_1\cdots dx_m$ is the Euclidean $m$-dimensional volume element and $\Sigma=\{\m{x}\in\R^m:\fhi(\m{x})=0\}$ is such that the gradient $\pa_{\m{x}}[\fhi]$ does not vanish on $\Sigma$. In this way, one may see the Lebesgue integral $\int_{\Sigma}f \, dS$ 
as a functional depending on an equation, namely $\fhi(\m{x})=0$. This approach has proven to be useful when defining integration over smooth hypersurfaces in superspace, see \cite{Guz_Somm5, MR2539324}.

The aim of the present paper is to tackle several problems concerning invariant integration over manifolds of arbitrary co-dimension $k<m$. Let us briefly discuss the problems and tasks that will be considered in this paper.

{As we have already mentioned, higher co-dimensional extensions of formula (\ref{HorFor}) have been used in random matrix theory in order to compute Haar integrals over some Lie groups. Our first goal is to summarize these ideas for the general case of implicitly defined submanifolds in $\R^m$.} 
Given a smooth $(m-k)$-dimensional manifold $\Sigma\inc \R^m$ defined by means of $k$ equations $\fhi_1(\m{x})=\ldots=\fhi_k(\m{x})=0$, we describe the Lebesgue integral $\int_\Sigma\, dS$ in terms of the concentrated Dirac distribution $\del(\fhi_1)\ldots \del(\fhi_k)$ (see Theorem \ref{Fund-Teo}).

This idea allows to replace general group (or coset) integrals with Haar measure by the action of delta distributions. 
{As a first application, we will use this method to compute integrals over the Stiefel manifolds $\textup{SO}(m)/\textup{SO}(m-k)$.} In \cite{MR3395222},  Pizzetti type formulas for these integrals were obtained. These formulas express the Stiefel integrals as series expansions of rotation invariant operators. A different proof for the Pizzetti formula in the case $\textup{SO}(m)/\textup{SO}(m-2)$ was derived in \cite{MR3591150}. It turns out that the distributional approach to invariant integration allows for an alternative proof for these formulas. In Section \ref{StSubSect}, we provide a detailed account on this new proof which simplifies the computations and reasonings from \cite{MR3395222, MR3591150}.

Our next problem is to provide a distributional interpretation to oriented integration. To that end, we first describe the integral of any ($m-k$)-differential form ${\bf \al}$ over $\Sigma$ in terms of the concentrated delta distribution as (see Theorem \ref{Teo2}) 
 \[\int_\Sigma \bal = \int_{\R^m} \del(\fhi_1)\ldots \del(\fhi_k) \, d\fhi_1\ldots d\fhi_k \, \bal.\]
 Next, we consider the case ${\bf \al}=f\Psi_{m-k}$ where $f$ is a smooth function on $\Sigma$ and $\Psi_{m-k}$ is the {\it oriented $(m-k)$-dimensional surface measure}, see e.g. \cite{MR1012510, MR3706179}. In this case, the above formula will provide a purely distributional expression for the oriented integral $\int_\Sigma f\Psi_{m-k}$, see Proposition \ref{OrFormul}.

This distributional approach can be applied to obtain a Cauchy type formula for the tangential Dirac operator on an $(m-k)$-surface. In the case where $k=0$, the classical Cauchy formula for the $m$-dimensional Dirac operator $\pa_{\m{x}}$ reads as (see e.g.\ \cite{MR1169463})
\begin{equation}\label{ClassCau1}
\int_{\pa C} F \Psi_{m-1} G = \int_{C} \left((F \pa_{\m{x}}) G + F (\pa_{\m{x}} G) \right)\, dV,
\end{equation}
where $F,G$ are smooth Clifford-algebra-valued functions on $C$ and $C\inc\R^m$ is a compact set  with smooth boundary $\pa C$. Extensions of this formula to any co-dimension $1\leq k<m$ were derived (by means of differential forms) as consequences of Stokes' theorem, see e.g. \cite{MR1012510, MR3706179}. As we will show in Section \ref{CauTTDO}, the Cauchy formula for the tangential Dirac operator can also be obtained from our distributional perspective, i.e.\ not using Stokes' theorem. We will provide a detailed description of both methods and show how they are linked together.

Summarizing, we deal with the following problems:
  \begin{itemize}
\item {\bf P1:} extend formula (\ref{HorFor}) to a general co-dimension $k$
\item {\bf P2:} prove Pizzetti's formulas for Stiefel manifolds from a distributional perspective
\item {\bf P3:} give a distributional interpretation to invariant oriented integration
\item {\bf P4:} prove a Cauchy type theorem for the tangential Dirac operator using distributions.
\end{itemize}

The notion of delta distribution concentrated on a manifold of lower dimension plays an important role in solving these problems. In this paper, we use the definition provided by Gel'fand and Shilov in \cite[Ch.~3]{MR0166596}. Their approach allows to exploit the link between generalized functions and integration of differential forms. This approach is also useful for derivatives of distributions concentrated on a manifold, which naturally arise  when dealing with Dirac distributions on supermanifolds. This shall be considered in future work.

{The paper is organized as follows. In Section 2, we provide some basic preliminaries on Clifford algebras and differential forms.  In Section \ref{Sec3ad}, we describe non-oriented integration over general manifolds in terms of distributions. To that end, we first recall some concepts related with the so-called Dirac distribution concentrated on a manifold. Next, we show how the Lebesgue integral on a smooth ($m-k$)-surface can be written as a particular action of this distribution, thus solving problem {\bf P1}. In Section \ref{StSubSect}, we apply this approach to easily compute integrals over Stiefel manifolds, providing a solution to {\bf P2}.  A more detailed account on integration over $\textup{SO}(m)/\textup{SO}(m-2)$ is given in Appendix A. In Section \ref{Sect4}, we study oriented integration in terms of distributions, which solves {\bf P3}. Finally, in Section \ref{CauTTDO}, we apply this approach in order to obtain a Cauchy formula for the tangential Dirac operator, hence solving {\bf P4}.}

\section{Preliminaries}
{In this section we fix some notation and recall some necessary concepts about Clifford algebras and differential forms that will be required in the sequel.}

The  real Clifford algebra $\mathbb{R}_m$ is the real associative algebra with generators $e_1,...,e_m$ satisfying the defining relations $e_je_{\el}+e_{\el}e_j=-2\delta_{j\el}$, for $j,\el=1,...,m$, where $\delta_{j\el}$ is the well-known Kronecker symbol. Every element $a\in \mathbb{R}_m$ can be written in the form $a=\sum_{A\inc M}a_Ae_A$, where $a_A\in \R$, 
$M:=\{1,\ldots, m\}$ and for any multi-index $A=\{j_1, \ldots, j_k\}\subseteq M$ with $j_1< \ldots < j_k$ we put $e_A=e_{j_1}\cdots e_{j_k}$ and $|A|=k$. Every $a\in\mathbb{R}_m$ admits a multivector decomposition 
\[a=\sum_{k=0}^m [a]_k, \hspace{.5cm} \mbox{ where } \hspace{.5cm} [a]_k=\sum_{|A|=k}a_A e_A.\]
Here $[\cdot]_k:\R_m\fd \R_m^{(k)}$ denotes the canonical projection of $\R_m$ onto the space of $k$-vectors $\R_m^{(k)}=\textup{span}_\R\{e_A:|A|=k\}$ . Note that $\R_m^{(0)}=\R$ is the set of scalars, while the space of $1$-vectors $\R_m^{(1)}$ is isomorphic to $\R^m$ via the identification $\m{v}=\sum_{j=1}^m v_j e_j\fd (v_1,\ldots, v_m)^T$.
When necessary, we shall use this interpretation of the $1$-vector $\m{v}\in\R_m^{(1)}$ as a column vector in $\R^m$.

The Clifford product of two vectors  $\m{v},\m{w}\in \mathbb{R}_m^{(1)}$ may be written as $\m{v}\,\m{w}=\m{v}\cdot \m{w}+ \m{v}\wedge\m{w}$
where 
\begin{align*}
\m{v}\cdot \m{w} &= \frac{1}{2}(\m{v}\,\m{w}+ \m{w}\,\m{v}) = -\sum_{j=1}^m v_jw_j,  & & \mbox{and} & \m{v}\wedge \m{w} &= \frac{1}{2}(\m{v}\,\m{w}- \m{w}\,\m{v}) = \sum_{j<\ell} (v_jw_\el-v_\el w_j) e_j e_\el,
\end{align*}
are the so-called dot and wedge products respectively. Note that $\m{v}\cdot \m{w}=-\langle\m{v},\m{w}\rangle \in \R$ where $\langle\cdot,\cdot\rangle$ denotes the Euclidean inner product in $\R^m$ while $ \m{v}\wedge \m{w}\in \R_m^{(2)}$. Furthermore, for $\m{v}_1, \ldots, \m{v}_k\in \R_m^{(1)}$ we define the wedge (or Grassmann) product in terms of the Clifford product by
\[\m{v}_1\wedge \cdots \wedge \m{v}_k =  \frac{1}{k!} \sum_{\pi\in\textup{Sym}(k)}\textup{sgn}(\pi)\, \m{v}_{\pi(1)} \cdots \m{v}_{\pi(k)} \in \R_m^{(k)}. \]
Among the most important properties of the wedge product we have
\begin{itemize}
\item  $\un{v}_{\pi(1)} \wedge \ldots \wedge \un{v}_{\pi(k)}= \text{sgn}(\pi) \un{v}_1 \wedge \ldots \wedge \un{v}_k$;
\item $\left[ \un{v}_1  \cdots \un{v}_k \right]_k = \un{v}_1 \wedge \ldots \wedge \un{v}_k $;
\item if $\{\un{v}_1, \ldots, \un{v}_k\}$ is a set of orthogonal vectors, then $\un{v}_1 \wedge \ldots \wedge \un{v}_k = \un{v}_1  \cdots \un{v}_k$;
\item $\un{v}_1 \wedge \ldots \wedge \un{v}_k=0$ if and only if  the vectors $\un{v}_1, \ldots, \un{v}_k$ are linearly dependent.
\end{itemize}
We shall also make use of the so-called {\it Gram matrix} of the vectors $\m{v}_1, \ldots, \m{v}_k\in \R^m$, which is defined as
\[G(\m{v}_1, \ldots, \m{v}_k)=M^T M= \left(\begin{array}{cccc} \langle\m{v}_1, \m{v}_1\rangle &  \langle\m{v}_1, \m{v}_2\rangle & \ldots & \langle\m{v}_1, \m{v}_k\rangle \\[+.1cm]  
\langle\m{v}_2, \m{v}_1\rangle &  \langle\m{v}_2, \m{v}_2\rangle & \ldots & \langle\m{v}_2, \m{v}_k\rangle \\[+.1cm]  
\vdots & \vdots& \ddots & \vdots\\[+.1cm]
 \langle\m{v}_k, \m{v}_1\rangle &  \langle\m{v}_k, \m{v}_2\rangle & \ldots & \langle\m{v}_k, \m{v}_k\rangle \end{array}\right),\]
 where $M=(\m{v}_1| \m{v}_2| \ldots| \m{v}_k)$ is the matrix whose columns are given by the vectors $\m{v}_j$ while $M^T=\left(\hspace{-.1cm}\begin{array}{c} \m{v}_1^T\\ \vdots \\ \m{v}_k^T\end{array} \hspace{-.1cm}\right)$ is the transpose of $M$. The {\it Gram determinant} is the determinant of $G(\m{v}_1, \ldots, \m{v}_k)$ and can be expressed in terms of the wedge product of vectors by
 \begin{equation}\label{Gram_Det}
 \det G(\m{v}_1, \ldots, \m{v}_k) = \|  \un{v}_1 \wedge \ldots \wedge \un{v}_k\|^2,
 \end{equation}
 where $\|\cdot\|$ denotes the Euclidean norm in $\R_m$ defined by $\|a\|^2=\sum_{A\subseteq M} a_A^2$ for  $a\in \R_m$. This determinant coincides with the square of the volume of the parallelepiped spanned by the vectors $\m{v}_1, \ldots, \m{v}_k$.

For $\m{v}\in\R^{(1)}_m$ and $a\in\R^{(k)}_m$ we set $\m{v}a=[\m{v}a]_{k-1}+[\m{v}a]_{k+1}=\m{v}\cdot a+\m{v}\wedge a$,
where 
\begin{align*}
\m{v}\cdot a&=[\m{v}a]_{k-1}=\frac{1}{2}(\m{v} a-(-1)^ka\m{v}), & & \mbox{and} & \m{v}\wedge a&=[\m{v}a]_{k+1}=\frac{1}{2}(\m{v} a+(-1)^ka\m{v}). 
\end{align*}
Furthermore, for $a\in\R^{(k)}_m$ and $b\in\R^{(\ell)}_m$, $k\leq \el$, we have
\begin{equation}\label{Mult_Decomp}
ab=[ab]_{\el-k}+[ab]_{\el-k+2}+\cdots+[ab]_{\el+k-2}+ [ab]_{\el+k}.
\end{equation}
We thus define the dot and wedge product of multivectors as $a\cdot b = [ab]_{\el-k}$ and $a\wedge b = [ab]_{\el+k}$ respectively. {They satisfy the following properties, see e.g.\ \cite{MR2077085,MR1169463}.}
\begin{pro}\label{ClifProp}
Let $\m{v}, \m{v}_1, \ldots, \m{v}_k\in \R_m^{(1)}$, $a=\sum_{j=0}^k [a]_j$ and $b\in\R_m^{(\el)}$ with $k \leq \el$. Then
\begin{itemize}
\item[$i)$] $\m{v}\cdot (b \, e_M)= (\m{v}\wedge b) e_M$;
\item[$ii)$] $[ab]_{\el-k}=\left[[a]_k b\right]_{\el-k}$ and $[ab]_{\el+k}=\left[[a]_k b\right]_{\el+k}$;
\item[$iii)$] $\left(\m{v}_1\wedge \cdots \wedge \m{v}_k\right)\cdot b=\left[\m{v}_1 \cdots \m{v}_k \, b\right]_{\el-k}=\m{v}_1 \cdot \left(\m{v}_2\cdot\left(\cdots\left(\m{v}_k\cdot b\right)\ldots\right)\right)$;
\item[$iv)$] $\left(\m{v}_1\wedge \cdots \wedge \m{v}_k\right)\wedge b=\left[\m{v}_1 \cdots \m{v}_k \, b\right]_{\el+k}=\m{v}_1 \wedge \left(\m{v}_2\wedge\left(\cdots\left(\m{v}_k\wedge b\right)\ldots\right)\right)$.
\end{itemize}
\end{pro}

The variable $(x_1, \ldots, x_m)$ in $\R^m$ is usually identified with the vector variable $\m{x}=\sum_{j=1}^m x_je_j$. In general, we consider $\R_m$-valued functions of the vector variable $\m{x}$, i.e.\ functions of the form $f(\m{x})=\sum_{A\subseteq M} f_A(\m{x}) e_A$ where the $f_A$'s are $\R$-valued functions. In that way, given a function space $\mathcal{F}=\C^k(\R^m), C^\infty(\Om)$, etc. we obtain the corresponding space of Clifford-valued functions  $\mathcal{F}\otimes \R_{m}$.

Throughout this paper we will use the so-called Dirac operator (or gradient) defined as
\[
\partial_{\m{x}}=e_1\partial_{x_1}+\cdots+e_m\partial_{x_m}.
\]
Null solutions of the Dirac operator are called {\it monogenic functions}.
The monogenic function theory constitutes a natural and successful extension of classical complex analysis to higher dimensions. As the Dirac operator
factorizes the Laplace operator:
\[\Del_{\m{x}}=-\pa_{\m{x}}^2=\sum_{j=1}^m \pa_{x_j}^2,\]
monogenicity also constitutes a refinement of harmonicity. Standard references on this setting are \cite{MR1130821, MR697564, MR1169463}.

We now recall some basic ideas on the theory of differential forms. For a more complete treatment see e.g.\  \cite{MR0166596, MR3706179, MR2723362} and the references therein. We denote by $\Lambda_k(C^\infty(\Om))$ the space of differential forms of $k$-th degree (or $k$-forms) on the open set $\Om\subseteq \R^m$.  Given the coordinate system $(x_1, \ldots, x_m)$, elements in $\Lambda_k(C^\infty(\Om))$ can be written as $\bal= \sum_{|A|=k} \al_A(\m{x})\, dx_A$ where $\al_A(\m{x})\in C^\infty(\Om)$ and $dx_A=dx_{j_1}\cdots dx_{j_k}$ for $A=\{j_1, \ldots, j_k\}$ with $1\leq j_1< \ldots < j_k\leq m$. Here the coordinate differentials $dx_j$'s satisfy the anticommutation rule $dx_j\, dx_k= -dx_k\, dx_j$. The exterior derivative $d: \Lambda_k(C^\infty(\Om)) \fd \Lambda_{k+1}(C^\infty(\Om))$ can be written in these coordinates as $d=\sum_{j=1}^m dx_j  \pa_{x_j}$. 

Orientability and integration over manifolds are of crucial importance in this paper. Let $V\inc\Om$ be a neighborhood of dimension $k\leq m$ defined as the image of a diffeomorphism $\m{x}(\cdot): V^\prime\inc\R^k\fd V$ where $\m{u}=(u_1,\ldots, u_k) \mapsto \m{x}(u_1,\ldots, u_k)$. The local coordinate system $\m{u}=(u_1,\ldots, u_k)$ is said to define an orientation in $V$. The same orientation is defined by any other local coordinate system $\m{v}=(v_1, \ldots, v_k)$ as long as the corresponding Jacobian $J\hspace{-.05cm}\left(\stackanchor{\m{u}}{\m{v}}\right)$ is positive.
It is also possible to give $V$ the opposite orientation by considering any local coordinate system $\m{w}=(w_1, \ldots, w_k)$ such that $J\hspace{-.05cm}\left(\stackanchor{\m{u}}{\m{w}}\right)<0$.

Let $\bal(\m{x})\in \Lambda_k(C^\infty(\Om))$ with compact support on $V$, i.e.\ $\textup{supp}\,\bal \cap V$ is compact. Then the integral of $\bal$ over $V$ equipped with the orientation given by $u_1,\ldots, u_k$ is defined by
\begin{equation}\label{IntDiffForm}
\int_{V} \bal := \int_{V^\prime} \sum_{|A|=k} \al_A(\m{x}(\m{u}))\, J\hspace{-.05cm}\left(\stackanchor{\m{x}_A}{\m{u}}\right)\, du_1\cdots du_{k},
\end{equation}
where $J\hspace{-.05cm}\left(\stackanchor{\m{x}_A}{\m{u}}\right)=J\hspace{-.05cm}\left(\stackanchor{x_{j_1}, \ldots, x_{j_k}}{u_{1}, \ldots, u_{k}}\right)$ is the Jacobian that appears automatically from the chain rule.
The integral $\int_{V} \bal$ is defined uniquely up to a sign, which is determined by the orientation of $V$.

In general, a $k$-dimensional manifold cannot be fully parametrized by an open subset of $\R^k$, but such a parametrization is always possible locally. A smooth manifold $\Sigma$ is called {\it orientable} if in a neighborhood of any point of $\Sigma$ it is possible to define the orientation consistently, i.e.\ in such a way that the coordinates in intersecting neighborhoods define the same orientation in the intersection. It is possible to integrate compactly supported differential forms over general smooth orientable manifolds. This integral can be defined as a finite sum of integrals (\ref{IntDiffForm}) over local parametrizations. We refer the reader to \cite[Ch.\ 6 ]{MR2723362}  for a detailed account on this procedure.

%

{\section{Concentrated delta distribution and non-oriented integration}\label{Sec3ad}
In this section we describe general non-oriented integration over submanifolds embedded in $\R^m$ in terms of the Dirac distribution. To that end, we first recall some basic notions related to the delta distribution concentrated on manifolds of dimension less than $m$ imbedded in $\R^m$ ($m>1$), see \cite[Ch.~3]{MR0166596} for more details. This distribution plays an essential role in computing integrals over implicitly defined surfaces, as it will be shown in the next sections. 
\subsection{The delta distribution concentrated on manifolds}
}

{Let us consider  an $(m-k)$-surface $\Sigma\inc \R^m$  defined by means of $k$  equations of the form}
\[\fhi_1(x_1, \ldots, x_m)=0, \hspace{.7cm} \fhi_2(x_1, \ldots, x_m)=0, \;\;\ldots,  \;\; \fhi_k(x_1, \ldots, x_m)=0,\]
where the so-called {\it defining phase functions} $\fhi_1, \ldots, \fhi_k\in C^\infty(\R^m)$ are {\it independent}, i.e.\ 
\[\pa_{\m{x}}[\fhi_1]\wedge \ldots\ \wedge \pa_{\m{x}}[\fhi_k]\neq 0\;\;\; \mbox{ on } \;\;\; \Sigma.\]
The previous condition means that, at any point of $\Sigma$, there is a $k$-blade orthogonal to $\Sigma$ and therefore a ($m-k$)-dimensional tangent plane. In particular, it can be proved that the following properties hold as well.
\begin{itemize}
\item The vectors $\pa_{\m{x}}[\fhi_1], \ldots, \pa_{\m{x}}[\fhi_k]$ are linearly independent on every point of $\Sigma$.
\item The matrix $\left(\begin{array}{ccc} \pa_{x_1}[\fhi_1] & \ldots & \pa_{x_m}[\fhi_1]\\ \vdots & \ddots & \vdots \\ \pa_{x_1}[\fhi_k] & \ldots & \pa_{x_m}[\fhi_k]\\\end{array}\right)$ has rank $k$ on every point of $\Sigma$.
\item $d\fhi_1\ldots d\fhi_k\neq 0$ on $\Sigma$.
\item In an $m$-dimensional neighborhood $U$ of any point of $\Sigma$, there exists a $C^\infty$-local coordinate system  in which  the first k coordinates are $u_1=\fhi_1$, $\ldots$, $u_k=\fhi_k$, and the remaining $u_{k+1}, \ldots, u_m$ can be choosen so that $J\hspace{-.1cm}\left(\stackanchor{\m{x}}{\m{u}}\right)>0$.
\end{itemize}
\begin{remark}\label{IndProp}
The last property implies that $\Sigma$ is an oriented $(m-k)$-surface. Indeed, this property shows how to introduce local coordinate systems $u_{k+1}, \ldots, u_m$ in the neighborhood of any point of $\Sigma$ with a consistent orientation. {This way of choosing the coordinate system $u_{k+1}, \ldots, u_m$ determines what we shall call the orientation of $\Sigma$ defined by the ordered $k$-tuple $(\fhi_1,\ldots, \fhi_k)$.}
\end{remark}
{In virtue of the last property above, for any test function $\phi \in C^\infty(\R^m)$ with support in $U$ one has}
\[
\int \del(\fhi_1)\cdots \del(\fhi_k)\phi(\m{x})\, dV=  \int \del(u_1)\cdots \del(u_k) \, \psi(\m{u}) \, du_1 \cdots du_m,
\]
where $\psi(\m{u})= \phi(\m{x}(\m{u}))J\hspace{-.1cm}\left(\stackanchor{\m{x}}{\m{u}}\right)$. It is thus natural to define $\del(\fhi_1)\cdots \del(\fhi_k)$ as 
\begin{equation}\label{del(P)_Def1}
(\del(\fhi_1)\ldots \del(\fhi_k), \phi ) = \int \del(\fhi_1)\cdots \del(\fhi_k)\phi(\m{x})\, dV=\int \psi(0,\ldots, 0, u_{k+1},\ldots, u_m) \, du_{k+1} \cdots du_m.
\end{equation}
This last integral is taken over the surface $\Sigma$, which is why the generalized function $\del(\fhi_1)\ldots \del(\fhi_k)$ is said to be concentrated on this surface. 

The definition given in formula (\ref{del(P)_Def1}) can be rewritten in terms of differential forms. Observe that, in the neighborhood $U$, the map $(u_{k+1},\ldots, u_m) \mapsto \m{x}(0,\ldots, 0, u_{k+1},\ldots, u_m)$ provides a parametrization for the surface $\Sigma$. Thus the integral in equation (\ref{del(P)_Def1}) can be written as $\int_{\Sigma\cap U} \phi \bomega$ where $\bomega = J\hspace{-.1cm}\left(\stackanchor{\m{x}}{\m{u}}\right) du_{k+1} \ldots du_m$ is an $(m-k)$-differential form uniquely determined on the surface  $\Sigma$ by the relation (see \cite[Ch.~3]{MR0166596})
\begin{equation}\label{wk}
d\fhi_1 \ldots d\fhi_k \, \bomega=dV.
\end{equation}
Finally,  the generalized function $\del(\fhi_1)\ldots \del(\fhi_k)$ can be redefined by 
\begin{equation}\label{delk}
(\del(\fhi_1)\ldots \del(\fhi_k), \phi ) :=  \int_{\R^m} \del(\fhi_1)\ldots \del(\fhi_k)\phi\, dV =\int_{\Sigma} \phi \, \bomega,
\end{equation}
where the orientation of $\Sigma$ is defined by the ordered $k$-tuple $(\fhi_1, \ldots, \fhi_k)$, see Remark \ref{IndProp}.

\subsection{Distributions and non-oriented integration over $(m-k)$-surfaces}\label{Sect3}
The goal of this section is to describe non-oriented integration of functions over the $(m-k)$-dimensional surface $\Sigma=\{\m{x}\in\R^m: \fhi_1(\m{x})=\ldots=\fhi_k(\m{x})=0\}$ in terms of the generalized function $\del(\fhi_1)\ldots \del(\fhi_k)$  defined in (\ref{delk}). In order to do so, we state first the following classic result from linear algebra. As in the previous section, we abuse of the notation $\m{v}\in \R_m^{(1)}$ to denote the corresponding column vector in $\R^m$.
\begin{lem}\label{LA1}
Let $M,M^{-1}\in\textup{GL}(m)$ be written in terms of their row and column vectors respectively, i.e.
{\small\[M=\left(\hspace{-.1cm}\begin{array}{c} \m{v}_1^T\\ \vdots \\ \m{v}_m^T\end{array} \hspace{-.1cm}\right), \;\;\;\; \mbox{ and } \;\;\;\; M^{-1}=(\m{w}_1|  \ldots| \m{w}_m).\]}
If the basis vectors $\m{v}_1, \ldots, \m{v}_m\in \R^m$ satisfy $\langle \m{v}_j, \m{v}_{k+\el}\rangle =0$ for $j=1, \ldots, k$ and $\el=1,\ldots, m-k$,
then the following properties hold
\begin{itemize}
\item[$i)$] $\displaystyle \langle \m{w}_j, \m{w}_{k+\el}\rangle =0,\;\;\;\;$  $\displaystyle j=1, \ldots, k,\;\;\;$  $\displaystyle \el=1,\ldots, m-k$,
\item[$ii)$] $|\det M| = \|\m{v}_1\wedge\ldots \wedge \m{v}_k\| \|\m{v}_{k+1}\wedge\ldots \wedge \m{v}_{m}\|$,
\item[$iii)$]  $\|\m{v}_1\wedge\ldots \wedge \m{v}_k\|  \|\m{w}_1\wedge\ldots \wedge \m{w}_k\|=1,\;\;$   $\|\m{v}_{k+1}\wedge\ldots \wedge \m{v}_m\|  \|\m{w}_{k+1}\wedge\ldots \wedge \m{w}_m\|=1$.
\end{itemize}
\end{lem}
\begin{remark}
Geometrically, property $ii)$ states that the volume of a parallelepiped formed by two  orthogonal frames equals the product of the volumes of the parallelepipeds corresponding to each frame.
\end{remark}

{ We also need the following result from differential geometry (see e.g.\ \cite{MR2954043}) which allows to locally split coordinate systems in $\Sigma$ into tangent and normal directions. In combination with the previous result, this leads to a factorization of the corresponding Jacobian that will be needed in the sequel.}
\begin{lem}\label{OrtSys}
Let $\Sigma\inc \R^m$ be a ($m-k$)-surface defined by means of the independent phase functions  $\fhi_1, \ldots, \fhi_k\in C^\infty(\R^m)$ as before. Then, in an $m$-dimensional neighborhood $U$ of any point of $\Sigma$, we can choose a $C^\infty$-coordinate system $u_1=\fhi_1$, $\ldots$, $u_k=\fhi_k$, $u_{k+1}, \ldots, u_m$ such that  $J\hspace{-.1cm}\left(\stackanchor{\m{x}}{\m{u}}\right)>0$ and
\begin{equation}\label{L2St}
\left\langle \pa_{\m{x}}[\fhi_j],\pa_{\m{x}}[u_{k+\el}]\right\rangle =0, \;\;\; \mbox{ on }\;\;\;\; U\cap\Sigma,
\end{equation}
for $j=1, \ldots, k$ and $\el=1,\ldots, m-k$.
\end{lem}

The combined use of the lemmas \ref{LA1} and \ref{OrtSys} allows to prove the following result. As before,
we consider an open region $\Om\inc \R^m$.
\begin{teo}\label{Fund-Teo}
Let $\Sigma\inc \Om$ be a ($m-k$)-surface defined by means of the independent phase functions  $\fhi_1, \ldots, \fhi_k\in C^\infty(\R^m)$. Then for any function $f \in C^\infty (\Om)$, with $\textup{supp}\, f \cap \Sigma$ compact, we have
\[\int_{\R^m} \del(\fhi_1)\ldots \del(\fhi_k) f \, dV  = \int_\Sigma \frac{f}{\left\|\pa_{\m{x}}[\fhi_1]\wedge \ldots\ \wedge \pa_{\m{x}}[\fhi_k]\right\|} \, dS,\]
where $dS$ is the $(m-k)$-dimensional Euclidean surface measure on $\Sigma$. In consequence, the following formula for non-oriented integration over $\Sigma$ holds
\begin{equation}\label{NO_Int_For}
\int_\Sigma f\, dS= \int_{\R^m} \del(\fhi_1)\ldots \del(\fhi_k) \left\|\pa_{\m{x}}[\fhi_1]\wedge \ldots\ \wedge \pa_{\m{x}}[\fhi_k]\right\|f \, dV. 
\end{equation}
\end{teo}
\begin{remark}
Theorem \ref{Fund-Teo} is a generalization to higher codimensions of Theorem 6.1.5 in \cite{MR1996773}.
\end{remark}
\pf
Lemma \ref{OrtSys} allows us to choose, in an $m$-dimensional neighborhood $U$ of a fixed but arbitrary point of $\Sigma$, a smooth coordinate system  $u_1=\fhi_1$, $\ldots$, $u_k=\fhi_k$, $u_{k+1}, \ldots, u_m$ with $J\hspace{-.1cm}\left(\stackanchor{\m{x}}{\m{u}}\right)>0$ and such that the orthogonality condition (\ref{L2St}) holds. This means that the coordinate axes of $u_{k+1}, \ldots, u_m$ are orthogonal to the ones of $\fhi_1,\ldots, \fhi_k$.
This coordinate system automatically provides a parametrization for the surface $U\cap \Sigma$ by means of the map (defined in some open set of $\R^{m-k}$) given by
\[\m{x}(\cdot): (u_{k+1},\ldots, u_m) \fd \m{x}(0, \ldots, 0, u_{k+1},\ldots, u_m).\]
Then the $(m-k)$-dimensional volume element $dS$ on $U\cap \Sigma$ is given by $dS = \sqrt{\det G} \; du_{k+1}\ldots du_m$, 
where $G=G\left(\pa_{u_{k+1}}[\m{x}], \ldots, \pa_{u_{m}}[\m{x}]\right)$ is the Gram matrix of the vectors $\pa_{u_{k+1}}[\m{x}], \ldots, \pa_{u_{m}}[\m{x}]$. Thus, in virtue of  (\ref{Gram_Det}), we have 
\[
dS=\left\|\pa_{u_{k+1}}[\m{x}]\wedge\ldots \wedge \pa_{u_{m}}[\m{x}]\right\|\; du_{k+1}\ldots du_m.
\]
\noindent On the other hand, let $\bomega$  be the $(m-k)$-differential form defined in (\ref{wk}) associated to $\del(\fhi_1)\ldots \del(\fhi_k)$. We know from (\ref{wk}) that $\bomega = J\hspace{-.1cm}\left(\stackanchor{\m{x}}{\m{u}}\right) du_{k+1} \ldots du_m$. We also recall that
\begin{align*}
J\hspace{-.1cm}\left(\stackanchor{\m{x}}{\m{u}}\right) &= \det \left(\pa_{u_{1}}[\m{x}] \big |  \ldots \big | \pa_{u_{m}}[\m{x}] \right),
& &\mbox{ and } &
J\hspace{-.1cm}\left(\stackanchor{\m{u}}{\m{x}}\right)&= \det \left(\hspace{-.1cm}\begin{array}{c} \pa_{\m{x}}[u_1]^T\\ \vdots \\ \pa_{\m{x}}[u_m]^T\end{array} \hspace{-.1cm}\right),
\end{align*}
where the vectors $\pa_{u_{j}}[\m{x}]=(\pa_{u_{j}}[x_1],\ldots, \pa_{u_{j}}[x_m])^T$ and $\pa_{\m{x}}[u_j]^T=(\pa_{x_{1}}[u_j],\ldots, \pa_{x_m}[u_j])$ represent column and row vectors respectively. Using the orthogonality condition (\ref{L2St}) and Lemma \ref{LA1} we obtain 
\begin{align*}\label{Jacob_Gram}
J\hspace{-.1cm}\left(\stackanchor{\m{x}}{\m{u}}\right) &= J\hspace{-.1cm}\left(\stackanchor{\m{u}}{\m{x}}\right)^{-1} =   \left\|\pa_{\m{x}}[u_1]\wedge\ldots \wedge \pa_{\m{x}}[u_k]\right\|^{-1} \left\|\pa_{\m{x}}[u_{k+1}]\wedge\ldots \wedge \pa_{\m{x}}[u_m]\right\|^{-1}= \frac{\left\|\pa_{u_{k+1}}[\m{x}]\wedge\ldots \wedge \pa_{u_{m}}[\m{x}]\right\|}{ \left\|\pa_{\m{x}}[u_1]\wedge\ldots \wedge \pa_{\m{x}}[u_k]\right\|}.
\end{align*}
Finally, for any $\phi\in C_0^\infty(U)$ we obtain
\begin{align*}
\int_{\R^m} \del(\fhi_1)\ldots \del(\fhi_k)\, \phi \, dV  &= \int_{U\cap \Sigma} \phi \,\bomega = \int_{U\cap \Sigma}  J\hspace{-.1cm}\left(\stackanchor{\m{x}}{\m{u}}\right) \phi \, du_{k+1} \ldots du_m \\
&= \int_{U\cap \Sigma} \phi \,  \frac{\left\|\pa_{u_{k+1}}[\m{x}]\wedge\ldots \wedge \pa_{u_{m}}[\m{x}]\right\| }{ \left\|\pa_{\m{x}}[u_1]\wedge\ldots \wedge \pa_{\m{x}}[u_k]\right\|}  \, du_{k+1} \ldots du_m \\
&= \int_{U\cap \Sigma} \frac{\phi }{\left\|\pa_{\m{x}}[\fhi_1]\wedge \ldots\ \wedge \pa_{\m{x}}[\fhi_k]\right\|} \, dS,
\end{align*}
which proves the theorem. $\hfill\square$

Non-oriented integration over $(m-k)$-surfaces, as described in formula (\ref{NO_Int_For}), requires only a set of $k$ equations defining the surface instead of a parametrization of it. This provides a useful approach for computing integrals over surfaces in which the parametrization constitutes a difficult issue. In particular, it allows to integrate over some compact Lie groups and coset spaces by replacing the Haar measure by the action of delta distributions. This will be illustrated in the next section where we integrate over Stiefel manifolds. In addition, this approach provides a suitable integration method over supermanifolds, which are mostly defined in an algebraic way. For more details on the case of co-dimension $k=1$, we refer the reader to \cite{Guz_Somm5, MR2539324}. The case of higher co-dimensions in superspace shall be regarded in future works.

As it is expected, the generalized function $\del(\fhi_1)\ldots \del(\fhi_k) \left\|\pa_{\m{x}}[\fhi_1]\wedge \ldots\ \wedge \pa_{\m{x}}[\fhi_k]\right\|$ does not change when we change the equations describing $\Sigma$. In order to study how $\del(\fhi_1)\ldots \del(\fhi_k)$  and $\pa_{\m{x}}[\fhi_1]\wedge \ldots\ \wedge \pa_{\m{x}}[\fhi_k]$ change, let us transform the equations $\fhi_1=\ldots=\fhi_k=0$ to $\psi_1=\ldots=\psi_k=0$ where 
\[\psi_\el(\m{x})= \sum_{j=1}^k \al_{\el,j} (\m{x}) \fhi_j(\m{x}), \hspace{1cm} \el=1,\ldots, k.\]
Here the functions $\al_{\el,j}\in C^\infty(\R^m)$ are such that the matrix they form is nonsingular, i.e.\ $\det\{\al_{\el,j}\}\neq 0$ for every $\m{x}\in \R^m$. Obviously both sets of equations define the same manifold $\Sigma$. 
If $\det\{\al_{\el,j}\} > 0$,  the ordered $k$-tuple $(\psi_1, \ldots, \psi_k)$ defines the same orientation of $\Sigma$ as $(\fhi_1, \ldots, \fhi_k)$. Otherwise, these $k$-tuples define opposite orientations, see Remark \ref{IndProp}.
 
The new differential form $\onda{\bomega}$ associated to $(\psi_1,\ldots, \psi_k)$ is given by $d\psi_1\ldots d\psi_k\, \onda{\bomega}=dV$. On the surface $\Sigma$, the differentials $d\psi_\el$ can be written as $d\psi_\el=\sum_{j=1}^k \left(\al_{\el,j}  \, d\fhi_j + \fhi_j \, d\al_{\el,j}\right)=\sum_{j=1}^k \al_{\el,j}  \, d\fhi_j$.
Then
\[dV=d\psi_1\ldots d\psi_k\, \onda{\bomega}= \left(\sum \al_{1,j}  \, d\fhi_j\right) \ldots \left(\sum \al_{k,j}  \, d\fhi_j\right) \onda{\bomega} = \det\{\al_{\el,j}\} \, d\fhi_1\ldots d\fhi_k \, \onda{\bomega},\]
which yields $\onda{\bomega} = \frac{1}{\det\{\al_{\el,j}\}}\, \bomega$.
Hence the relation between $\del(\psi_1)\ldots \del(\psi_k)$  and $\del(\fhi_1)\ldots \del(\fhi_k)$ is given by
\begin{equation*}
(\del(\psi_1)\ldots \del(\psi_k), \phi ) =\int_{\Sigma} \phi \, \onda{\bomega}  = \sgn\left(\det\{\al_{\el,j}\}\right) \int_{\Sigma}  \frac{\phi}{\det\{\al_{\el,j}\}} \, \bomega = \left(\del(\fhi_1)\ldots \del(\fhi_k),  \frac{\phi}{\|\det\{\al_{\el,j}\}\|} \right),
\end{equation*}
where the orientation of $\Sigma$ in the first integral is defined by $(\psi_1,\ldots, \psi_k)$, while the orientation in the second integral is defined by $(\fhi_1, \ldots, \fhi_k)$. This justifies the appearance of the absolute value $\|\det\{\al_{\el,j}\}\|= \sgn\left(\det\{\al_{\el,j}\}\right) \det\{\al_{\el,j}\}$, where $\sgn\left(\det\{\al_{\el,j}\}\right)=\pm 1$ is the sign of $\det\{\al_{\el,j}\}$. We then obtain
\[\del(\psi_1)\ldots \del(\psi_k) = \frac{\del(\fhi_1)\ldots \del(\fhi_k)}{\|\det\{\al_{\el,j}\}\|}.\]
On the surface $\Sigma$ we also have $\pa_{\m{x}}[\psi_\el]= \sum_{j=1}^k \al_{\el,j}  \, \pa_{\m{x}}[\fhi_j]$. Therefore,
\[\pa_{\m{x}}[\psi_1]\wedge \ldots\ \wedge \pa_{\m{x}}[\psi_k] = \left(\sum_{j=1}^k \al_{1,j}  \, \pa_{\m{x}}[\fhi_j]\right)\wedge \ldots \wedge \left(\sum_{j=1}^k \al_{k,j}  \, \pa_{\m{x}}[\fhi_j]\right)= \det\{\al_{\el,j}\} \, \pa_{\m{x}}[\fhi_1]\wedge \ldots\ \wedge \pa_{\m{x}}[\fhi_k]. \]
In this way, we have obtained the following result.
\begin{pro}\label{InvPro}
Let $\Sigma\inc \R^m$ be a ($m-k$)-surface defined by means of the independent phase functions  $\fhi_1, \ldots, \fhi_k\in C^\infty(\R^m)$ and let $\psi_\el= \sum_{j=1}^k \al_{\el,j}  \fhi_j$,  $\el=1,\ldots, k$, be new functions such that $\al_{\el,j}\in C^\infty(\R^m)$ and $\det\{\al_{\el,j}\}\neq 0$ for every $\m{x}\in \R^m$. We then have 
\[\del(\psi_1)\ldots \del(\psi_k) \, \pa_{\m{x}}[\psi_1]\wedge \ldots\ \wedge \pa_{\m{x}}[\psi_k]= \sgn\left(\det\{\al_{\el,j}\}\right) \del(\fhi_1)\ldots \del(\fhi_k) \,\pa_{\m{x}}[\fhi_1]\wedge \ldots\ \wedge \pa_{\m{x}}[\fhi_k]\]
and 
\[\del(\psi_1)\ldots \del(\psi_k) \, \left\|\pa_{\m{x}}[\psi_1]\wedge \ldots\ \wedge \pa_{\m{x}}[\psi_k]\right\|=  \del(\fhi_1)\ldots \del(\fhi_k) \,\left\|\pa_{\m{x}}[\fhi_1]\wedge \ldots\ \wedge \pa_{\m{x}}[\fhi_k]\right\|. \]
\end{pro}

\section{Pizzetti formulae for Stiefel manifolds}\label{StSubSect}
In this section we direct our attention towards some applications of the above description of integration over $(m-k)$-surfaces in terms of the action of delta distributions. In particular, we use formula (\ref{NO_Int_For}) to derive an alternative proof of the so-called Pizzetti formula for real Stiefel Manifolds $\textup{St}(m,k)$ of any order $0<k\leq m$, see \cite{MR3395222}.

The Stiefel manifold $\textup{St}(m,k)$ is the set of all orthonormal $k$-frames in $\mathbb{R}^{m}$ ($m>1$), i.e.\  the set of ordered $k$-tuples of orthonormal vectors in $\R^m$. If we write a $k$-frame as a matrix $M$ of $k$ column vectors in $\R^m$ we have
\[\textup{St}(m,k) := \{M\in \R^{m\times k}: M^TM=\mathbbm{1}_k\},\]
which is equivalent to
\[\textup{St}(m,k) := \{(\m{x}_1, \ldots, \m{x}_k)\in \left(\R^m\right)^k: \langle\m{x}_j, \m{x}_\el\rangle=\del_{j,\el}, \; j,\el=1,\ldots, k\}. \]
Thus $\textup{St}(m,k)$ is a compact $\left(mk-\frac{k(k+1)}{2}\right)$-dimensional manifold embedded in $\R^{mk}$. The orthogonal group O$(m)$ acts transitively on $\textup{St}(m,k)$ and, when $k<m$, the special orthogonal group SO$(m)$ also acts transitively on this manifold. Then it follows that $\textup{St}(m,k)$ can be seen as a homogeneous space
\begin{align*}
\textup{St}(m,k) &\iso \textup{O}(m)/\textup{O}(m-k)\\
&\iso \textup{SO}(m)/\textup{SO}(m-k) \;\;\;\;\; (\mbox{if } \;\; k<m).
\end{align*}

\noindent When $k = m$ or $k=m-1$, it is easily seen that  $\textup{St}(m,k)$ is diffeomorphic to the corresponding classical groups $\textup{St}(m,m)\iso\textup{O}(m)$ and $\textup{St}(m,m-1)\iso\textup{SO}(m)$ respectively. 

Now let us examine integration over Stiefel manifolds of any order.

\begin{paragraph}{Integration over $\textup{St}(m,1)=\Sa^{m-1}$.}
Clearly, the Stiefel manifold of first order coincides with the unit sphere $\Sa^{m-1}$ in $\R^m$ since a $1$-frame is nothing but a unit vector. Pizzetti's formula provides a method to compute integrals over $\Sa^{m-1}$ by acting with a certain power series of the Laplacian operator on the integrand, see \cite{Pizz}. For any polynomial $P:\R^m\fd \C$, this formula reads as 
\begin{equation}\label{PizSph}
\int_{\Sa^{m-1}} P(\m{x}) \; {d S_{\m{x}}} =  \sum_{k=0}^{\infty}  \frac{2 \pi^{m/2}}{2^{2k} k!\Gamma (k+m/2)} \left(\Delta^k_{\m{x}} P \right)(0) = \Phi_m(\Del_{\m{x}})[P] \bigg|_{\m{x}=0},
\end{equation}
where $\displaystyle \Phi_m(z)=(2\pi)^{\frac{m}{2}} \frac{J_{\frac{m}{2}-1}(iz^{1/2})}{(iz^{1/2})^{\frac{m}{2}-1}}$ and $J_\nu$ is the Bessel function of first kind. This power series, defined by the renormalized Bessel function $\Phi_m(z)$, involves only natural powers of the argument $z$.
\end{paragraph}
%
 
 \begin{paragraph}{Integration over $\textup{St}(m,2)$.}
The Stiefel manifold of order 2 is defined as  \[\textup{St}(m,2) = \{(\m{x},\m{y})\in \left(\R^m\right)^2: \|\m{x}\|=\|\m{y}\|=1, \langle \m{x}, \m{y}\rangle=0\}.\]
The integral over  $\textup{St}(m,2)$ can be interpreted as the integral over the $(m-1)$-dimensional sphere $\Sa^{m-1}$ with respect to $\m{x}$ and the integral over the $(m-2)$-dimensional subsphere $\Sa^{m-2}_\perp$ with respect to $\m{y}$, that is perpendicular to $\m{x}\in\Sa^{m-1}$. Then for any polynomial $P(\m{x},\m{y}):\R^m\times \R^m \fd \C$ one has
\begin{align}\label{GenIntForm}
\int_{\textup{St}(m,2)}P(\m{x}, \m{y})  \, dS_{\m{x},\m{y}} = \int_{\Sa^{m-1}} \left(\int_{\Sa^{m-2}_\perp} P(\m{x}, \m{y}) \, dS_{\m{y}}\right) \, dS_{\m{x}}.
\end{align}
Let us compute first the innermost integral for any fixed $\m{x}\in \Sa^{m-1}$. Observe that the $(m-2)$-surface $\Sa_\perp^{m-2}$ is defined by the pair of smooth functions $\fhi_1(\m{y})=\|\m{y}\|-1$ and $\fhi_2(\m{y})=\langle \m{x},\m{y}\rangle$. The gradients of these phase functions satisfy the following identity on  $\Sa_\perp^{m-2}$
\[\left\| \pa_{\m{y}}[\fhi_1]\wedge \pa_{\m{y}}[\fhi_2]\right\|=\frac{\left\| \m{x}\wedge\m{y}\right\|}{\left\| \m{y}\right\|} = \frac{\left\| \m{x}\right\| \left\| \m{y}\right\|}{\left\| \m{y}\right\|}=\left\| \m{x}\right\|=1. \]
Thus, according to formula (\ref{NO_Int_For}), we have
\[ \int_{\Sa^{m-2}_\perp} P(\m{x}, \m{y}) \, dS_{\m{y}} = \int_{\R^m} \del(\|\m{y}\|-1) \del(\langle \m{x},\m{y}\rangle) \, P(\m{x}, \m{y}) \;dV_{\m{y}}.\]
\end{paragraph}
Let us consider now the coordinate transformation $\m{u}=M\m{y}$ where $M\in \textup{SO}(m)$ is a rotation matrix whose first row is given by the unit vector $\m{x}\in \Sa^{m-1}$, i.e.\  the first component of $\m{u}$ is $u_1=\langle\m{x},\m{y}\rangle$.
Under this transformation, the above integral transforms into
\begin{align*}
\int_{\Sa^{m-2}_\perp} P(\m{x}, \m{y}) \, dS_{\m{y}} &= \int_{\R^m} \del(\|\m{u}\|-1) \del(u_1) \, P(\m{x},M^{-1}\m{u}) \;du_1\ldots du_m \\
&=  \int_{\R^{m-1}} \del((u_2^2+\cdots+u_m^2)^{1/2}-1) \,P(\m{x},M^{-1}(0, u_2,\ldots, u_m)^T)  \;du_2\ldots du_m.
\end{align*}
%
The last expression coincides with the integral over the sphere $\Sa^{m-2}\inc \R^{m-1}$ with respect to the vector $(u_2, \ldots, u_m)$. Thus applying Pizzetti's formula (\ref{PizSph}) over this sphere yields
\begin{align}\label{FirstPizAp}
\int_{\Sa^{m-2}_\perp} P(\m{x}, \m{y}) \, dS_{\m{y}} &= \Phi_{m-1}(\Del_{\m{u}}-\pa_{u_1}^2) P(\m{x},M^{-1}\m{u})\big|_{\m{u}=0} \nonumber \\
&= \Phi_{m-1}(\Del_{\m{y}}-\langle \m{x},\pa_{\m{y}}\rangle^2) P(\m{x},\m{y})\big|_{\m{y}=0}
\end{align}
%
where we have used the identities $\Del_{\m{y}}=\Del_{\m{u}}$ and $\pa_{u_1}= \langle \m{x}, \pa_{\m{y}}\rangle= \sum_{j=1}^m x_j\pa_{y_j}$.
Substituting (\ref{FirstPizAp}) into (\ref{GenIntForm}) and applying  Pizzetti's formula for $\m{x}\in \Sa^{m-1}$ we obtain
\begin{equation}\label{PizStie2_1}
\int_{\textup{St}(m,2)}P(\m{x}, \m{y})  \, dS_{\m{x},\m{y}} = \Phi_{m}(\Del_{\m{x}}) \circ \Phi_{m-1}(\Del_{\m{y}}-\langle \m{x},\pa_{\m{y}}\rangle^2) P(\m{x},\m{y})\bigg|_{\m{x}=0 \atop \m{y}=0}.
\end{equation}
In this way, we have obtained a Pizzetti formula for $\textup{St}(m,2)$ by consecutively applying  Pizzetti's formula for the unit spheres in $\R^m$ and $\R^{m-1}$. The above formula can be stated in a more explicit way where the action of the vector variable $\m{x}$ is not present anymore, see 
\cite{MR3395222, MR3591150}. This connection is explicitly established in appendix \ref{PizS2}.

\begin{paragraph}{Integration over $\textup{St}(m,k)$.}
The same idea can be followed to integrate over the surface 
\[\textup{St}(m,k) := \{(\m{x}_1, \ldots, \m{x}_k)\in \left(\R^m\right)^k: \langle\m{x}_j, \m{x}_\el\rangle=\del_{j,\el}, \; j,\el=1,\ldots, k\}. \]
In this case, the integral of a polynomial $P(\m{x}_1, \ldots,\m{x}_k): \R^m\times \cdots\times \R^m\fd \C$ is given by
\begin{equation}\label{BigInt}
\int_{\St(m,k)} P(\m{x}_1, \ldots, \m{x}_k)\,  dS_{\m{x}_1, \ldots, \m{x}_k}= \int_{\Sa^{m-1}} \left( \int_{\Sa^{m-2}} \cdots \left(\int_{\Sa^{m-k}} P(\m{x}_1, \ldots, \m{x}_k)\,  dS_{\m{x}_k}\right) \ldots dS_{\m{x}_2}\right)dS_{\m{x}_1},
\end{equation}
where $\m{x}_j\in \Sa^{m-j}$ and $\Sa^{m-j}$ denotes the $(m-j)$-dimensional unit sphere in $\R^m$ that is perpendicular to $\m{x}_1,\ldots, \m{x}_{j-1}$. As in the previous case, each of these $k$ integrals can be computed by means of spherical Pizzetti's formulas. 

Let us illustrate this procedure by computing the innermost integral, provided that the orthonormal ($k-1$)-frame $(\m{x}_1, \ldots, \m{x}_{k-1})\in \textup{St}(m,k-1)$ has been fixed. The $(m-k)$-surface $\Sa^{m-k}$ is defined by the independent phase functions 
\[\fhi_1(\m{x}_k)=\|\m{x}_k\|-1, \;\; \fhi_2(\m{x}_k)=\langle \m{x}_1,\m{x}_k\rangle , \;\; \ldots, \;\; \fhi_k(\m{x}_k)=\langle \m{x}_{k-1},\m{x}_k\rangle,\]
whose gradients satisfy 
$\left\| \pa_{\m{y}}[\fhi_1]\wedge \cdots \wedge \pa_{\m{y}}[\fhi_k] \right\|=\frac{\| \m{x}_1\wedge\cdots \wedge \m{x}_k\|}{\| \m{x}_k\|} = \frac{\| \m{x}_1\| \cdots \| \m{x}_k\|}{\| \m{x}_k\|}=1$ on $\Sa^{m-k}$. Thus
\[\int_{\Sa^{m-k}} P(\m{x}_1, \ldots, \m{x}_k)\,  dS_{\m{x}_k} = \int_{\R^m} \del(\|\m{x}_k\|-1)  \del(\langle \m{x}_1,\m{x}_k\rangle) \cdots \del(\langle \m{x}_{k-1},\m{x}_k\rangle) \, P(\m{x}_1, \ldots, \m{x}_k)\;  dV_{\m{x}_k}.\]
As before, we consider $\m{u}=M\m{x}_k$ where the first $k-1$ rows of the matrix $M\in \textup{SO}(m)$ are given now by the orthonormal vectors $\m{x}_1, \ldots, \m{x}_{k-1}$. This means that the first $k-1$ components of the vector  $\m{u}$ are $u_j=\langle \m{x}_j,\m{x}_k\rangle$, $j=1,\ldots, k-1$. Effectuating this change of variables in the above integral we obtain
\begin{align*}
&\int_{\Sa^{m-k}} P(\m{x}_1, \ldots, \m{x}_k)\,  dS_{\m{x}_k} \\
&\;\;\;  = \int_{\R^m} \del(\|\m{u}\|-1)  \del(u_1) \cdots \del(u_{k-1}) \, P(\m{x}_1, \ldots, \m{x}_{k-1}, M^{-1}\m{u} )\;  du_1\ldots du_m\\
&\;\;\;  = \int_{\R^{m-k+1}} \del((u_k^2+\cdots+u_m^2)^{1/2}-1) P(\m{x}_1, \ldots, \m{x}_{k-1}, M^{-1}(0, \ldots, 0, u_k,\ldots, u_m)^T )\; du_k\ldots du_m,
\end{align*}
where the last integral is taken over the sphere $\Sa^{m-k}\inc \R^{m-k+1}$ with respect to the vector $(u_k, \ldots, u_m)$. Then Pizzetti's formula (\ref{PizSph}) yields
\begin{align*}
\int_{\Sa^{m-k}} P(\m{x}_1, \ldots, \m{x}_k)\,  dS_{\m{x}_k} &  =\Phi_{m-k+1} \left(\Del_{\m{u}}-\pa_{u_1}^2 - \ldots - \pa_{u_{k-1}}^2\right) P(\m{x}_1, \ldots, \m{x}_{k-1}, M^{-1}\m{u} ) \big|_{\m{u}=0}\\
& = \Phi_{m-k+1}\Big(\Del_{\m{x}_k}-\langle \m{x}_1,\pa_{\m{x}_k}\rangle^2 - \ldots - \langle \m{x}_{k-1},\pa_{\m{x}_k}\rangle^2 \Big) P(\m{x}_1, \ldots, \m{x}_{k} ) \big|_{\m{x}_k=0}.
\end{align*}
Repeating the same procedure for the other $(k-1)$ iterated integrals in (\ref{BigInt}) we obtain the following result in terms of the composition of $k$ {\it spherical} Pizzetti formulas in dimensions $m, m-1, \ldots, m-k+1$. 
\begin{teo}\label{PizT}{\bf [Pizzetti formula for $\St(m,k)$]}
For any polynomial $P(\m{x}_1, \ldots,\m{x}_k): \R^m\times \cdots\times \R^m\fd \C$ it holds that 
\[ \int_{\textup{St}(m,k)}P(\m{x}_1, \ldots,\m{x}_k)  \, dS_{\m{x}_1,\ldots ,\m{x}_k}
  =\prod_{j=1}^k \Phi_{m-j+1}\left(\Del_{\m{x}_j} - \sum_{\el=1}^{j-1} \langle \m{x}_\el,\pa_{\m{x}_j}\rangle^2\right) \,  [P]\bigg|_{\m{x}_1,\ldots, \m{x}_k =0}.
\]
\end{teo} 
\end{paragraph}

\section{Distributions and oriented integration over $(m-k)$-surfaces}\label{Sect4}
In this section we describe oriented integration over the $(m-k)$ surface $\Sigma$ in terms of the distribution $\del(\fhi_1)\ldots \del(\fhi_k)$. To that end, we start by studying integration of general differential forms in $\Lam_{m-k}(C^\infty(\Om))$ where $\Om\inc \R^m$ is an open set that contains $\Sigma$.
\begin{teo}\label{Teo2}
Let $\Sigma\inc \Om$ be a ($m-k$)-surface defined by means of the independent phase functions  $\fhi_1, \ldots, \fhi_k\in C^\infty(\R^m)$ and let $\bal\in\Lam_{m-k}(C^\infty(\Om))$ with $\textup{supp}\, \bal\cap \Sigma$ compact. Then
\[\int_\Sigma \bal = \int_{\R^m} \del(\fhi_1)\ldots \del(\fhi_k) \, d\fhi_1\ldots d\fhi_k \, \bal.\]
\end{teo}
\begin{remark}
Observe that $d\fhi_1\ldots d\fhi_k \, \bal$ is a differential form of maximum degree $m$, i.e.\ it can be written as $d\fhi_1\ldots d\fhi_k \, \bal= f \, dV$ where $f \in C^\infty (\Om)$ is such that $\textup{supp}\, f\cap \Sigma$ is compact. Then the integral in the right-hand side is well-defined in the sense of (\ref{delk}).
\end{remark}
\pf
It is enough to prove this result in an $m$-dimensional neighborhood $U$ of any point of $\Sigma$ and to assume that ${\bf \al}$ is compactly supported on $U\cap \Sigma$. Let us consider a local coordinate system in $U$ of the form $u_1=\fhi_1$, $\ldots$, $u_k=\fhi_k$, $u_{k+1}, \ldots, u_m$ such that $J\hspace{-.1cm}\left(\stackanchor{\m{x}}{\m{u}}\right)>0$, see Remark \ref{IndProp}.  
 When we write the differential form $\bal$ in this coordinate system we get
\begin{align*}
\bal &= \phi \, du_{k+1}\ldots du_m+ \sum_{j=1}^k \bbe_j \, d\fhi_j, 
\end{align*}
where $\phi\in C^\infty(U)$ has compact support on $U\cap \Sigma$ and $\bbe_j\in\Lam_{m-k-1}(C^\infty(U))$.
Therefore we can write $\int _{U\cap \Sigma} \bal = \int _{U\cap \Sigma}  \phi \, du_{k+1}\ldots du_m.$
Now we recall that the differential form $\bomega$ defined in (\ref{wk}) can be written as $\bomega = J\hspace{-.1cm}\left(\stackanchor{\m{x}}{\m{u}}\right)  du_{k+1}\ldots du_m $. From the definition of $\del(\fhi_1)\ldots \del(\fhi_k)$ in (\ref{delk}) we obtain
\[\int _{U\cap \Sigma} \bal=  \int _{U\cap \Sigma}  \phi \,  J\hspace{-.1cm}\left(\stackanchor{\m{x}}{\m{u}}\right)^{-1} \bomega = \int_{\R^m} \del(\fhi_1)\ldots \del(\fhi_k) \, \phi \,  J\hspace{-.1cm}\left(\stackanchor{\m{x}}{\m{u}}\right)^{-1}  dV.\]
Finally, since $dV= J\hspace{-.1cm}\left(\stackanchor{\m{x}}{\m{u}}\right)  d\fhi_1\ldots d\fhi_k \, du_{k+1}\ldots du_m$, we get
\[ \int _{U\cap \Sigma} \bal= \int_{\R^m} \del(\fhi_1)\ldots \del(\fhi_k) \, d\fhi_1\ldots d\fhi_k \, \left(\phi \, du_{k+1}\ldots du_m \right)=  \int_{\R^m} \del(\fhi_1)\ldots \del(\fhi_k) \, d\fhi_1\ldots d\fhi_k \, \bal,\]
which proves the theorem. $\hfill\square$

\begin{defi}
The oriented  $(m-k)$-surface element in $\R^m$  is defined by the differential form
\[\Psi_{m-k} = \sum_{|A|=k} (-1)^{\el(A)} \, e_A \, dx_{M\setminus A} = \sum_{|B|=m-k}  (-1)^{\el(M\setminus B)} \, e_{M\setminus B} \, dx_B, \]
where as before $M:=\{1,\ldots, m\}$, $A=\{j_1, \ldots, j_k\}\subseteq M$ with $j_1< \ldots < j_k$ and $\el(A) = (j_1-1)+ (j_2-2)+ \ldots + (j_k-k)$ is the integer satisfying
 \[dV=  (-1)^{\el(A)} dx_A dx_{M\setminus A}, \hspace{.5cm} \mbox{ and }  \hspace{.5cm} e_M= (-1)^{\el(A)} e_A e_{M\setminus A}.\]
\end{defi}
\begin{remark}\label{OSurElPartCas}
Observe that $\Psi_{m-k}$ is a $k$-vector valued $(m-k)$-differential form. {When integrating on the smooth manifold $\Sigma$, the factors $e_A$ in $\Psi_{m-k}$ allow to recover the $k$-blade that is normal to $\Sigma$.} In particular, $\Psi_{m}=dV$ and when $k=1$ we can write 
\[\Psi_{m-1}=\sum_{j=1}^m (-1)^{j-1} e_j \widehat{dx_j}= n(\m{x}) \, dS,\]
where $\widehat{dx_j}=dx_1\ldots dx_{j-1}dx_{j+1}\ldots dx_m$, $n(\m{x})$ is the normal vector to $\Sigma$ at the point $\m{x}\in \Sigma$ and $dS$ is the Euclidean surface measure on $\Sigma$.
\end{remark}
 The element $\Psi_{m-k}$ is used in the following definition of oriented integration over $(m-k)$-surfaces.
  \begin{defi}\label{OrInt}
 The oriented integral of a function $f\in C^\infty(\Om)$ over the $(m-k)$-surface $\Sigma$ is given by
 \[\int_\Sigma f\Psi_{m-k} = \int_{\R^m} \del(\fhi_1)\ldots \del(\fhi_k) \, d\fhi_1\ldots d\fhi_k \, f\Psi_{m-k}. \]
 \end{defi}
 The second integral in the above definition directly follows from Theorem \ref{Teo2}. This expression can be rewritten in a purely distributional way on account of the following result. 
 \begin{pro}\label{OrFormul}
The following identity holds
\begin{equation}\label{OrDifform}
d\fhi_1\ldots d\fhi_k \, \Psi_{m-k}=  \pa_{\m{x}}[\fhi_1]\wedge \ldots\ \wedge \pa_{\m{x}}[\fhi_k] \, dV.
\end{equation}
In consequence, the oriented integral in Definition \ref{OrInt} can be written as
\begin{align}\label{IntDistOrDifform}
\int_\Sigma f\Psi_{m-k} &=  \int_{\R^m} \del(\fhi_1)\ldots \del(\fhi_k) \, \pa_{\m{x}}[\fhi_1]\wedge \ldots\ \wedge \pa_{\m{x}}[\fhi_k] \, f \, dV = \int_{\Sigma} \frac{\pa_{\m{x}}[\fhi_1]\wedge \ldots\ \wedge \pa_{\m{x}}[\fhi_k]}{ \|\pa_{\m{x}}[\fhi_1]\wedge \ldots\ \wedge \pa_{\m{x}}[\fhi_k] \|} \, f \, dS.
\end{align}
\end{pro}
\begin{remark}
The second equality in formula (\ref{IntDistOrDifform}) states that the oriented integral of $f$ over $\Sigma$ is the Lebesgue integral of the function $f$ multiplied by the unit $k$-blade  $\frac{\pa_{\m{x}}[\fhi_1]\wedge \ldots\ \wedge \pa_{\m{x}}[\fhi_k]}{ \|\pa_{\m{x}}[\fhi_1]\wedge \ldots\ \wedge \pa_{\m{x}}[\fhi_k] \|}$ that is normal to $\Sigma$. This constitutes a generalization to higher co-dimensions of the case where $k=1$, see Remark \ref{OSurElPartCas}.
\end{remark}
\pf
We compute
\begin{align*}
d\fhi_1\ldots d\fhi_k \, \Psi_{m-k} &=  \left(\sum_{j=1}^m \pa_{x_j}[\fhi_1] dx_j\right) \ldots  \left(\sum_{j=1}^m \pa_{x_j}[\fhi_k] dx_j\right)  \Psi_{m-k} \\
&=\left( \sum_{|A|=k} \left(\pa_{\m{x}}[\fhi_1]\wedge \ldots\ \wedge \pa_{\m{x}}[\fhi_k]\right)_A dx_A\right) \left( \sum_{|A|=k} (-1)^{\el(A)} \, e_A \, dx_{M\setminus A}\right) \\
&=  \sum_{|A|=k}   \left(\pa_{\m{x}}[\fhi_1]\wedge \ldots\ \wedge \pa_{\m{x}}[\fhi_k]\right)_A \, e_A \, (-1)^{\el(A)}  dx_A dx_{M\setminus A} \\
&= \pa_{\m{x}}[\fhi_1]\wedge \ldots\ \wedge \pa_{\m{x}}[\fhi_k] \, dV,
\end{align*}
which proves formula (\ref{OrDifform}). Formula (\ref{IntDistOrDifform}) is a direct consequence of Theorem \ref{Fund-Teo}.
$\hfill\square$

In a similar way to the non-oriented case, formula  (\ref{IntDistOrDifform}) describes oriented integration over ($m-k$)-surfaces  by means only of delta distributions of the $k$ defining phase functions. Moreover, Proposition \ref{InvPro} ensures that the generalized function $ \del(\fhi_1)\ldots \del(\fhi_k) \, \pa_{\m{x}}[\fhi_1]\wedge \ldots\ \wedge \pa_{\m{x}}[\fhi_k]$ does not depend on the choice of the defining $k$-tuple $(\fhi_1, \ldots, \fhi_k)$ as long as the orientation of $\Sigma$ is preserved.
%
%

In the works \cite{MR3706179, MR1012510}, the ($m-k$)-surface element was given in terms of the the vector differential $d\m{x}=\sum_{j=1}^m e_j dx_j$. In fact, one has the following relation. 
\begin{lem}\label{Lem4}
The following identity holds
\[\frac{(d\m{x})^{m-k}}{(m-k)!} = (-1)^{\frac{k(k+1)}{2}} \Psi_{m-k}\, e_M.\]
\end{lem}
\pf The equalities $e_M=(-1)^{\ell(A)} e_A e_{M\setminus A}$ and $e_A^2=(-1)^{\frac{k(k+1)}{2}} $ yield $e_{M\setminus A}= (-1)^{\el(A)+\frac{k(k+1)}{2}} e_A e_M$. Thus,
\[\frac{(d\m{x})^{m-k}}{(m-k)!} =  \sum_{|A|=k} e_{M\setminus A} dx_{M\setminus A} = (-1)^{\frac{k(k+1)}{2}} \left( \sum_{|A|=k} (-1)^{\el(A)} e_{ A} dx_{M\setminus A} \right) e_M= (-1)^{\frac{k(k+1)}{2}} \Psi_{m-k} e_M.\]
$\hfill\square$

Proposition \ref{OrFormul} can also be proven by examining the above relation between $\Psi_{m-k}$ and $d\m{x}$. In order to do so, we need the following lemma.
\begin{lem}\label{Lem5}
Let $\fhi\in C^\infty(\R^m)$. Then
\[\pa_{\m{x}}[\fhi] \cdot \frac{(d\m{x})^{k}}{k!} = - d\fhi \,\frac{(d\m{x})^{k-1}}{(k-1)!}.\]
\end{lem}
\pf
Observe that $\displaystyle\pa_{\m{x}}[\fhi] \cdot \frac{(d\m{x})^{k}}{k!} = \sum_{1\le j\le m\atop |A|=k} \pa_{x_j}[\fhi] dx_A \, (e_j\cdot e_A) = \sum_{j=1}^m  \pa_{x_j}[\fhi] \sum_{A\ni j} dx_A\, e_je_A$. Then
\[\pa_{\m{x}}[\fhi] \cdot \frac{(d\m{x})^{k}}{k!} =-  \sum_{j=1}^m  \pa_{x_j}[\fhi] dx_j \sum_{A\ni j} dx_{A\setminus\{j\}}\, e_{A\setminus\{j\}}= -d\fhi \sum_{|B|=k-1} dx_B e_B = - d\fhi \,\frac{(d\m{x})^{k-1}}{(k-1)!}.\]
$\hfill\square$

We are now able to obtain the following version of Proposition \ref{OrFormul}.
\begin{pro}\label{Lem6}
Let $\fhi_1, \ldots, \fhi_k\in C^\infty(\R^m)$ be independent phase functions. Then
\[ \pa_{\m{x}}[\fhi_1]\wedge \ldots\ \wedge \pa_{\m{x}}[\fhi_k] \, \frac{(d\m{x})^m}{m!} = (-1)^{\frac{k(k+1)}{2}} d\fhi_1\ldots d\fhi_k \, \frac{(d\m{x})^{m-k}}{(m-k)!}.\]
\end{pro}
\pf 
Since $\frac{(d\m{x})^m}{m!}= dV e_M$, the product  $\pa_{\m{x}}[\fhi_1]\wedge \ldots\ \wedge \pa_{\m{x}}[\fhi_k] \, \frac{(d\m{x})^m}{m!} $ gives an $(m-k)$-vector valued differential form. Then, by Proposition \ref{ClifProp} $iii)$ and Lemma \ref{Lem5}, we get
\begin{align*}
 \pa_{\m{x}}[\fhi_1]\wedge \ldots\ \wedge \pa_{\m{x}}[\fhi_k] \cdot \frac{(d\m{x})^m}{m!} &= \pa_{\m{x}}[\fhi_1] \cdot\left( \cdots \left( \pa_{\m{x}}[\fhi_k] \cdot \frac{(d\m{x})^m}{m!}\right)\ldots\right) \\
 &= -  \pa_{\m{x}}[\fhi_1] \cdot\left( \cdots \pa_{\m{x}}[\fhi_{k-1}] \cdot \left( d\fhi_k \, \frac{(d\m{x})^{m-1}}{(m-1)!}\right)\ldots\right)\\
 &\ldots\\
& =(-1)^k d\fhi_k d\fhi_{k-1}\ldots d\fhi_1 \, \frac{(d\m{x})^{m-k}}{(m-k)!}\\
 &= (-1)^{\frac{k(k+1)}{2}} d\fhi_1\ldots d\fhi_k \, \frac{(d\m{x})^{m-k}}{(m-k)!}.
\end{align*}
$\hfill\square$
\begin{remark}
As previously announced, Proposition \ref{Lem6} is equivalent to formula (\ref{OrDifform}). Indeed, by Lemma \ref{Lem4} we obtain 
\[(-1)^{\frac{k(k+1)}{2}} d\fhi_1\ldots d\fhi_k \, \frac{(d\m{x})^{m-k}}{(m-k)!}= d\fhi_1\ldots d\fhi_k \, \Psi_{m-k}\, e_M.\]
Thus $d\fhi_1\ldots d\fhi_k \, \Psi_{m-k}\, e_M=  \pa_{\m{x}}[\fhi_1]\wedge \ldots\ \wedge \pa_{\m{x}}[\fhi_k] \frac{(d\m{x})^m}{m!}= \pa_{\m{x}}[\fhi_1]\wedge \ldots\ \wedge \pa_{\m{x}}[\fhi_k]  \, dV\, e_M$.
\end{remark}

\section{Cauchy formulae for the tangential Dirac operator}\label{CauTTDO}
As an application of the previous approach, in this section we prove a Cauchy-type theorem for the tangential Dirac operator on an $(m-k)$-surface. As before, we consider an open set $\Om\inc \R^m$. 

In the case where $k=0$, the classical Cauchy formula (\ref{ClassCau1}) 
can be rewritten using distributions as follows. Let $C=\{\m{x}\in \R^m: \fhi(\m{x})\leq 0\}$ where the phase function $\fhi\in C^\infty(\R^m)$ is such that  $\pa_{\m{x}}[\fhi]\neq 0$ on $\pa C = \{\m{x}\in \R^m: \fhi(\m{x})= 0\}$. Then the surface integral $\int_{\pa C}$  can be written in distributional terms using Proposition \ref{OrFormul}, while the domain integral $\int_{C}$ can be written as the action of the Heaviside distribution 
\[H(-\fhi)=\begin{cases} 1 & \fhi\leq 0,\\ 0 & \fhi>0. \end{cases}\]
This way we can rewrite the  Cauchy formula (\ref{ClassCau1}) as
\begin{equation}\label{ClassCau}
\int_{\R^m} F \, \del(\fhi) \pa_{\m{x}}[\fhi] \, G \;dV=\int_{\R^m} H(-\fhi) \left((F \pa_{\m{x}}) G + F (\pa_{\m{x}} G) \right)\, dV.
\end{equation}

Our goal is to obtain a version of formula (\ref{ClassCau}) in any co-dimension $1\leq k < m$. To that end, we need to integrate over a compact $(m-k)$-surface namely $\Sigma \cap C$ and over its boundary $\Sigma \cap \pa C$. As before, $\Sigma\inc \Om$ is a ($m-k$)-surface defined by means of independent phase functions  $\fhi_1, \ldots, \fhi_k\in C^\infty(\R^m)$, and we shall also assume that  $\pa_{\m{x}}[\fhi]\wedge\pa_{\m{x}}[\fhi_1]\wedge \ldots\ \wedge \pa_{\m{x}}[\fhi_k]\neq 0$ on $\Sigma\cap \pa C$. We will illustrate two different methods for proving a Cauchy formula for the tangential Dirac operator on $\Sigma \cap C$. The first method is purely distributional, while the second one uses differential forms and Stokes' theorem. 

\subsection{The distributional approach}
We start by noticing that the distribution $H(-\fhi)\, \del(\fhi_1)\ldots \del(\fhi_k) \,    \pa_{\m{x}}[\fhi_1]\wedge \ldots\ \wedge \pa_{\m{x}}[\fhi_k]$ has compact support. Then for any $\R$-valued function $f\in C^\infty(\Om)$ we have
\begin{equation}\label{MainDistCauc}
\int_{\R^m} \pa_{\m{x}} \left[ f \, H(-\fhi)\, \del(\fhi_1)\ldots \del(\fhi_k) \,   \pa_{\m{x}}[\fhi_1]\wedge \ldots\ \wedge \pa_{\m{x}}[\fhi_k] \right] \, dV=0.
\end{equation}
The Cauchy formula follows after working out the integrand in the previous formula and taking the $(k+1)$-vector part. In order to do so, we need first the following technical lemma.
\begin{lem}\label{TechDelt}
The $(k+1)$-vector part of $\pa_{\m{x}}\left[\del(\fhi_1)\ldots \del(\fhi_k) \,   \pa_{\m{x}}[\fhi_1]\wedge \ldots\ \wedge \pa_{\m{x}}[\fhi_k]\right]$ vanishes.
\end{lem}
\pf
Observe that 
\begin{align*}
\pa_{\m{x}} & \left[ \del(\fhi_1)\ldots \del(\fhi_k) \,   \pa_{\m{x}}[\fhi_1]\wedge \ldots\ \wedge \pa_{\m{x}}[\fhi_k]\right] \\
&= \del(\fhi_1)\ldots \del(\fhi_k) \;    \pa_{\m{x}}\left[\pa_{\m{x}}[\fhi_1]\wedge \ldots\ \wedge \pa_{\m{x}}[\fhi_k]\right]+ \pa_{\m{x}}\left[\del(\fhi_1)\ldots \del(\fhi_k) \right] \,  \pa_{\m{x}}[\fhi_1]\wedge \ldots\ \wedge \pa_{\m{x}}[\fhi_k]  \\
&= \del(\fhi_1)\ldots \del(\fhi_k) \;    \pa_{\m{x}}\left[\pa_{\m{x}}[\fhi_1]\wedge \ldots\ \wedge \pa_{\m{x}}[\fhi_k]\right] 
+ \sum_{j=1}^k \del(\fhi_1)\ldots \del'(\fhi_j) \cdots \del(\fhi_k) \, \pa_{\m{x}}[\fhi_j] \pa_{\m{x}}[\fhi_1]\wedge \ldots\ \wedge \pa_{\m{x}}[\fhi_k]
 \end{align*}
where $\left[\pa_{\m{x}}[\fhi_j] \pa_{\m{x}}[\fhi_1]\wedge \ldots\ \wedge \pa_{\m{x}}[\fhi_k]\right]_{k+1}= \pa_{\m{x}}[\fhi_j]\wedge \pa_{\m{x}}[\fhi_1]\wedge \ldots\ \wedge \pa_{\m{x}}[\fhi_k]=0$. Thus the $(k+1)$-vector part of $\pa_{\m{x}}\left[\del(\fhi_1)\ldots \del(\fhi_k) \,   \pa_{\m{x}}[\fhi_1]\wedge \ldots\ \wedge \pa_{\m{x}}[\fhi_k]\right]$ is
\[ \del(\fhi_1)\ldots \del(\fhi_k) \;   \big[ \pa_{\m{x}}\left[\pa_{\m{x}}[\fhi_1]\wedge \ldots\ \wedge \pa_{\m{x}}[\fhi_k]\right]\big]_{k+1}.\]
Now we recall that the action of the Dirac operator on a product of $\R_m$-valued functions is given by $\pa_{\m{x}}[FG]=\pa_{\m{x}}[F] G + \overset{\circ}{\pa_{\m{x}}}[F\overset{\circ}{G}]$ where the overdot notation indicates the action of the partial derivatives, i.e.\ $\overset{\circ}{\pa_{\m{x}}}[F\overset{\circ}{G}]= \sum_{j=1}^m e_j F \pa_{x_j}[G]$. Using this idea we compute 
\begin{align*}
\big[ \pa_{\m{x}}\left[\pa_{\m{x}}[\fhi_1]\wedge \ldots\ \wedge \pa_{\m{x}}[\fhi_k]\right]\big]_{k+1} &= \sum_{j=1}^k
 \left[ \overset{\circ}{\pa_{\m{x}}} \left[\pa_{\m{x}}[\fhi_1]\wedge \ldots  \overset{\circ}{\pa_{\m{x}}[\fhi_j]}   \ldots \wedge \pa_{\m{x}}[\fhi_k]\right]\right]_{k+1} \\
 &= \sum_{j=1}^k (-1)^{j-1} \left[ \overset{\circ}{\pa_{\m{x}}} \left[ \overset{\circ}{\pa_{\m{x}}[\fhi_j]} \wedge \pa_{\m{x}}[\fhi_1]\wedge \ldots  \widehat{\pa_{\m{x}}[\fhi_j]}   \ldots \wedge \pa_{\m{x}}[\fhi_k] \right]\right]_{k+1},
 \end{align*}
 where $\widehat{z}$ means that the symbol $z$ is suppressed. Note that ${\pa_{\m{x}}[\fhi_j]} \wedge \pa_{\m{x}}[\fhi_1]\wedge \ldots  \widehat{\pa_{\m{x}}[\fhi_j]}   \ldots \wedge \pa_{\m{x}}[\fhi_k]$ is the $k$-vector part corresponding the Clifford product ${\pa_{\m{x}}[\fhi_j]} \pa_{\m{x}}[\fhi_1] \ldots  \widehat{\pa_{\m{x}}[\fhi_j]}   \ldots  \pa_{\m{x}}[\fhi_k]$. Thus, in virtue of Proposition \ref{ClifProp} $ii)$, we obtain 
 \begin{align*}
\big[ \pa_{\m{x}}\left[\pa_{\m{x}}[\fhi_1]\wedge \ldots\ \wedge \pa_{\m{x}}[\fhi_k]\right]\big]_{k+1} &=   \sum_{j=1}^k (-1)^{j-1} \left[ \overset{\circ}{\pa_{\m{x}}} \left[ \overset{\circ}{\pa_{\m{x}}[\fhi_j]} \pa_{\m{x}}[\fhi_1] \ldots  \widehat{\pa_{\m{x}}[\fhi_j]}   \ldots  \pa_{\m{x}}[\fhi_k] \right]\right]_{k+1} \\
&=  \sum_{j=1}^k (-1)^{j-1} \left[  \Del[\fhi_j] \, \pa_{\m{x}}[\fhi_1] \ldots  \widehat{\pa_{\m{x}}[\fhi_j]}   \ldots  \pa_{\m{x}}[\fhi_k] \right]_{k+1},
 \end{align*}
which implies that $\big[ \pa_{\m{x}}\left[\pa_{\m{x}}[\fhi_1]\wedge \ldots\ \wedge \pa_{\m{x}}[\fhi_k]\right]\big]_{k+1}=0$.
$\hfill\square$

Now we can proceed with our proof of a Cauchy formula. Working out the integrand in formula (\ref{MainDistCauc}), and taking into account that $\pa_{\m{x}}[H(-\fhi)]=- \del(\fhi) \pa_{\m{x}}[\fhi]$, we get
 \begin{align*}
 &\pa_{\m{x}}  \left[ f \, H(-\fhi)\, \del(\fhi_1)\ldots \del(\fhi_k) \,   \pa_{\m{x}}[\fhi_1]\wedge \ldots\ \wedge \pa_{\m{x}}[\fhi_k] \right]\\
&=  H(-\fhi)\, \del(\fhi_1)\ldots \del(\fhi_k) \,   \pa_{\m{x}}[f] \pa_{\m{x}}[\fhi_1]\wedge \ldots\ \wedge \pa_{\m{x}}[\fhi_k] - f \del(\fhi) \del(\fhi_1)\ldots \del(\fhi_k) \,   \pa_{\m{x}}[\fhi] \pa_{\m{x}}[\fhi_1]\wedge \ldots\ \wedge \pa_{\m{x}}[\fhi_k] \\
 &\phantom{=} +f \, H(-\fhi) \, \pa_{\m{x}}\left[\del(\fhi_1)\ldots \del(\fhi_k) \,   \pa_{\m{x}}[\fhi_1]\wedge \ldots\ \wedge \pa_{\m{x}}[\fhi_k]\right].
 \end{align*}
 Taking the $(k+1)$-vector part and applying Lemma \ref{TechDelt} we obtain the following Cauchy formula for the $\R$-valued function $f\in C^\infty(\Om)$,
\begin{multline}\label{For2}
\int_{\R^m} H(-\fhi)\, \del(\fhi_1)\ldots \del(\fhi_k) \,  \pa_{\m{x}}[f] \wedge  \pa_{\m{x}}[\fhi_1]\wedge \ldots\ \wedge \pa_{\m{x}}[\fhi_k]  \, dV \\=   \int_{\R^m}  \del(\fhi) \del(\fhi_1)\ldots \del(\fhi_k) \, \pa_{\m{x}}[\fhi]\wedge\pa_{\m{x}}[\fhi_1]\wedge \ldots\ \wedge \pa_{\m{x}}[\fhi_k]  \, f \, dV.
\end{multline}
This result can be easily extended to $\R_m$-valued functions by linearity. In particular, we obtain the following Cauchy formula.
\begin{teo}{\bf [Cauchy formula]}
For every $F,G\in C^\infty(\Om)\otimes \R_m$ one has
\begin{multline}\label{For3}
\int_{\R^m} H(-\fhi)\,\prod_{j=1}^k \del(\fhi_j) \left[ \overset{\circ}{F} \left( \overset{\circ}{\pa_{\m{x}}} \wedge  \pa_{\m{x}}[\fhi_1]\wedge \ldots\ \wedge \pa_{\m{x}}[\fhi_k]\right) \, G + F \left( \overset{\circ}{\pa_{\m{x}}} \wedge  \pa_{\m{x}}[\fhi_1]\wedge \ldots\ \wedge \pa_{\m{x}}[\fhi_k]\right)\overset{\circ}{G} \right]  \, dV \\
=  \int_{\R^m} \del(\fhi) \,\prod_{j=1}^k \del(\fhi_j) \, F\,  \pa_{\m{x}}[\fhi]\wedge\pa_{\m{x}}[\fhi_1]\wedge \ldots\ \wedge \pa_{\m{x}}[\fhi_k]  \, G \, dV,
\end{multline}
where the overdot notation indicates the action of the partial derivatives, i.e.\ if $F=\sum_{A}F_A e_A$ with $F_A\in C^\infty(\Om)$ then 
\[\overset{\circ}{F} \left( \overset{\circ}{\pa_{\m{x}}} \wedge  \pa_{\m{x}}[\fhi_1]\wedge \ldots\ \wedge \pa_{\m{x}}[\fhi_k]\right) = \sum_{A} e_A \, \pa_{\m{x}}[F_A] \wedge  \pa_{\m{x}}[\fhi_1]\wedge \ldots\ \wedge \pa_{\m{x}}[\fhi_k],
\] 
and 
\[\left( \overset{\circ}{\pa_{\m{x}}} \wedge  \pa_{\m{x}}[\fhi_1]\wedge \ldots\ \wedge \pa_{\m{x}}[\fhi_k]\right) \overset{\circ}{F}  = \sum_{A}  \pa_{\m{x}}[F_A] \wedge  \pa_{\m{x}}[\fhi_1]\wedge \ldots\ \wedge \pa_{\m{x}}[\fhi_k] \, e_A .
\] 
\end{teo}

Let us study more in detail the first order differential operator $ \pa_{\m{x}} \wedge  \pa_{\m{x}}[\fhi_1]\wedge \ldots\ \wedge \pa_{\m{x}}[\fhi_k]$, which is $(k+1)$-vector valued. In order to do so, we decompose the Dirac operator as 
\[\pa_{\m{x}}=\pa_{\m{x}\parallel} + \pa_{\m{x}\perp}  \]
where $\pa_{\m{x}\parallel}$ is the tangential part and $\pa_{\m{x}\perp}$ the normal part to the surface $\Sigma$. Given an arbitrary point $\m{w}\in \Sigma$, these operators can be explicitly written as follows. In a neighborhood of $\m{w}$, consider a $C^\infty$-coordinate system as in Lemma \ref{OrtSys}, i.e.\ 
\[u_1=\fhi_1, \ldots, u_k=\fhi_k, u_{k+1}, \ldots, u_m, \;\;\; \mbox{ with } \;\;\; \left\langle \pa_{\m{x}}[\fhi_j],\pa_{\m{x}}[u_{k+\el}]\right\rangle =0.\]
The space $N=\textup{span}_\R \{\pa_{\m{x}}[\fhi_1], \ldots, \pa_{\m{x}}[\fhi_k]\}$ is the $k$-dimensional plane orthogonal to $\Sigma$, while the space $T=\textup{span}_\R \{\pa_{\m{x}}[u_{k+1}], \ldots, \pa_{\m{x}}[u_m]\}$
is the $(m-k)$-dimensional plane tangent to $\Sigma$. Let $\{\m{\nu}_1, \ldots, \m{\nu}_k\}$ and $\{\m{\ve}_1, \ldots, \m{\ve}_{m-k}\}$ be orthonormal bases for the spaces $N$ and $T$ respectively.
We then have
\[\pa_{\m{x}\parallel}=\sum_{j=1}^{m-k} \m{\ve}_j \, \langle \m{\ve}_j , \pa_{\m{x}}\rangle, \hspace{.5cm} \mbox{ and } \hspace{.5cm} \pa_{\m{x}\perp}=\sum_{j=1}^{k} \m{\nu}_j \, \langle \m{\nu}_j , \pa_{\m{x}}\rangle.\]
It is clear that $\pa_{\m{x}\perp}\wedge \pa_{\m{x}}[\fhi_1]\wedge \ldots\ \wedge \pa_{\m{x}}[\fhi_k] =0$ and moreover,
\begin{equation}\label{TangDirOpMinus1}
\pa_{\m{x}} \wedge  \pa_{\m{x}}[\fhi_1]\wedge \ldots\ \wedge \pa_{\m{x}}[\fhi_k]= \pa_{\m{x}\parallel} \left(  \pa_{\m{x}}[\fhi_1]\wedge \ldots\ \wedge \pa_{\m{x}}[\fhi_k]\right) = (-1)^k  \left(  \pa_{\m{x}}[\fhi_1]\wedge \ldots\ \wedge \pa_{\m{x}}[\fhi_k]\right) \pa_{\m{x}\parallel}. 
\end{equation}
In this way, the $(k+1)$-vector valued operator  $ \pa_{\m{x}} \wedge  \pa_{\m{x}}[\fhi_1]\wedge \ldots\ \wedge \pa_{\m{x}}[\fhi_k]$ has been decomposed as the Clifford product of the $k$-vector $ \pa_{\m{x}}[\fhi_1]\wedge \ldots\ \wedge \pa_{\m{x}}[\fhi_k]$ with the tangential Dirac operator $\pa_{\m{x}\parallel}$. Thus formula (\ref{For3}) can be rewritten as
\begin{multline}\label{For3-1}
\int_{\R^m} H(-\fhi)\prod_{j=1}^k \del(\fhi_j) \left[ \left(F\pa_{\m{x}\parallel}\right) \,\pa_{\m{x}}[\fhi_1]\wedge \ldots\ \wedge \pa_{\m{x}}[\fhi_k] \, G + (-1)^k F\, \pa_{\m{x}}[\fhi_1]\wedge \ldots\ \wedge \pa_{\m{x}}[\fhi_k] \left(\pa_{\m{x}\parallel} G\right) \right]  \, dV \\
=  \int_{\R^m} \del(\fhi) \prod_{j=1}^k \del(\fhi_j) \, F\,  \pa_{\m{x}}[\fhi]\wedge\pa_{\m{x}}[\fhi_1]\wedge \ldots\ \wedge \pa_{\m{x}}[\fhi_k]  \, G \, dV.
\end{multline}

%

\subsection{The differential form approach}
The above technique for proving the Cauchy formula is purely distributional and does not involve any differential forms. Now we shall present another method that makes use of differential forms and Stokes' theorem. We consider the compact set $C\inc \Om$ and the $(m-k)$-surface $\Sigma\inc \Om$ as previously defined.

Let $\bal\in {\Lam}_{m-k-1}(C^\infty(\Om))$. From Stokes' theorem in the compact surface $\Sigma\cap C$ we obtain
\begin{equation}\label{StkesDfFo}
\int_{\Sigma\cap \pa C}  \bal = \int_{\Sigma\cap C}   d\bal,
\end{equation}
where the orientation in the boundary $\Sigma\cap \pa C$ is induced by the orientation of $\Sigma\cap C$. Since we consider the orientation of $\Sigma\cap C$ to be defined by the ordered $k$-tuple $(\fhi_1,\ldots, \fhi_k)$, the orientation of $\Sigma\cap \pa C$ is defined by $(\fhi_1,\ldots, \fhi_k, \fhi)$, see Remark \ref{IndProp}.

Cauchy's formula follows from working out formula (\ref{StkesDfFo}) when substituting $\bal=F\Psi_{m-k-1}G$ with $F,G\in C^\infty(\Om)\otimes \R_m$. To that end, we will need the following version of Lemma 4.
\begin{lem}\label{Lem5a}
The following operator identity holds in $C^\infty(\Om)$
\[\pa_{\m{x}} \wedge \Psi_{m-k}= (-1)^k  \, d \, \Psi_{m-k-1}.\]
\end{lem}
\pf
From Lemma \ref{Lem5} we know that $\pa_{\m{x}} \cdot \frac{(d\m{x})^{m-k}}{(m-k)!} = - d \,\frac{(d\m{x})^{m-k-1}}{(m-k-1)!}$.
By Lemma \ref{Lem4}, this equality can be rewritten as
\[\pa_{\m{x}} \cdot\left(\Psi_{m-k} \, e_M\right) = (-1)^k d \,\Psi_{m-k-1} \, e_M.\]
Finally, in virtue of Proposition \ref{ClifProp} $i)$, we obtain
\[\left(\pa_{\m{x}} \wedge \Psi_{m-k}\right) \, e_M= (-1)^k  \, d \, \Psi_{m-k-1} \, e_M,\]
which proves the Lemma. $\hfill\square$

For a given pair of functions $F,G\in C^\infty(\Om)\otimes \R_m$, the previous result yields 
\[(-1)^k d\left(F\Psi_{m-k-1} G\right)= \overset{\circ}{F} \overset{\circ}{\pa_{\m{x}}} \wedge \Psi_{m-k} \, G + F\, \overset{\circ}{\pa_{\m{x}}} \wedge \Psi_{m-k} \overset{\circ}{G}.\]
Therefore, by Stokes theorem we have 
\[ \int_{\Sigma\cap  C}  \left(\overset{\circ}{F} \overset{\circ}{\pa_{\m{x}}} \wedge \Psi_{m-k} \, G + F\, \overset{\circ}{\pa_{\m{x}}} \wedge \Psi_{m-k} \overset{\circ}{G}\right)= (-1)^k \int_{\Sigma\cap \pa C} F\Psi_{m-k-1} G.\]
From Proposition \ref{OrFormul} we know that $\Psi_{m-k}$ can be written as the multiplication of the Lebesgue measure on $\Sigma$ with the unit normal $k$-blade  $\frac{\pa_{\m{x}}[\fhi_1]\wedge \ldots\ \wedge \pa_{\m{x}}[\fhi_k]}{ \|\pa_{\m{x}}[\fhi_1]\wedge \ldots\ \wedge \pa_{\m{x}}[\fhi_k] \|}$. Thus it is readily seen that (see (\ref{TangDirOpMinus1}))
\[ {\pa_{\m{x}}} \wedge \Psi_{m-k}= \pa_{\m{x}\parallel}  \; \Psi_{m-k} = (-1)^k  \Psi_{m-k}\;  \pa_{\m{x}\parallel}.\]
Therefore,
\begin{equation}\label{DifFormCauchy}
 \int_{\Sigma\cap  C}  \left( \left(F\pa_{\m{x}\parallel}\right)  \Psi_{m-k} \, G+ (-1)^k F \, \Psi_{m-k} \left(\pa_{\m{x}\parallel}G\right)\right)= (-1)^k \int_{\Sigma\cap \pa C} F\Psi_{m-k-1} G.
 \end{equation}
The above formula exactly coincides with formula (\ref{For3-1}). Indeed, in virtue of Proposition \ref{OrFormul}, the left-hand side of (\ref{DifFormCauchy}) can be written as
\[\int_{\R^m} H(-\fhi)\prod_{j=1}^k \del(\fhi_j) \left[ \left(F\pa_{\m{x}\parallel}\right) \,\pa_{\m{x}}[\fhi_1]\wedge \ldots\ \wedge \pa_{\m{x}}[\fhi_k] \, G + (-1)^k F\, \pa_{\m{x}}[\fhi_1]\wedge \ldots\ \wedge \pa_{\m{x}}[\fhi_k] \left(\pa_{\m{x}\parallel} G\right) \right]  \, dV.\]
Now we recall that the orientation of $\Sigma\cap \pa C$ is defined by the ordered $(k+1)$-tuple $(\fhi_1, \ldots, \fhi_k,\fhi)$. Thus the right-hand side of  (\ref{DifFormCauchy}) coincides with
\[(-1)^k\int_{\R^m} F \del(\fhi_1)\ldots \del(\fhi_k)\del(\fhi)\, \pa_{\m{x}}[\fhi_1]\wedge \ldots\ \wedge \pa_{\m{x}}[\fhi_k]\wedge\pa_{\m{x}}[\fhi]\, G \, dV.\]
Finally, substituting these two expressions into (\ref{DifFormCauchy}) we obtain (\ref{For3-1}).
%

\section*{Acknowledgements}
The authors want to thank Hendrik de Bie and Michael Wutzig for their careful reading of the manuscript
and their very valuable suggestions. Alí Guzmán Adán is supported by a BOF-post-doctoral grant from Ghent University.

\appendix
\section{Further examining the Pizzetti formula for $\St(m,2)$}\label{PizS2}
The Pizzetti formula given in (\ref{PizStie2_1}) for $\St(m,2)$ can be written in terms of distributions as
\begin{align} \label{PrimFormS2}
\int_{\textup{St}(m,2)}P(\m{x}, \m{y})  \, dS_{\m{x},\m{y}} &= \left( \del(\m{x})\del(\m{y}) \,,\, \Phi_{m}(\Del_{\m{x}})  \Phi_{m-1}(\Del_{\m{y}}-\langle \m{x},\pa_{\m{y}}\rangle^2) [P] \right) \nonumber  \\
&= \left( \del(\m{y}) \,,\, \left(\Phi_{m}(\Del_{\m{x}})[\del(\m{x})] \, \Phi_{m-1}(\Del_{\m{y}}-\langle \m{x},\pa_{\m{y}}\rangle^2) \,,\, P \right) \right)
\end{align}
where $\del(\m{x})=\del(x_1)\cdots \del(x_m)$ and $\del(\m{y})=\del(y_1)\cdots \del(y_m)$ are the delta distributions for the vector variables $\m{x}$ and $\m{y}$ respectively. The innermost action with respect to $\m{x}$ can be expressed in a more explicit form depending only on differential operators with constant coefficients.
In order to do this we must examine the distribution 
\begin{equation}\label{Dist_inX}
 \Phi_{m}(\Del_{\m{x}})[\del(\m{x})] \, \Phi_{m-1}(\Del_{\m{y}}-\langle \m{x},\pa_{\m{y}}\rangle^2).
\end{equation}
\begin{lem}\label{FiscDual}
Let $R,Q \in \R[x_1, \ldots, x_m]$ be polynomials and $R(\pa_{\m{x}}),Q(\pa_{\m{x}}) \in \R[\pa_{x_1}, \ldots, \pa_{x_m}]$ be the corresponding Fischer duals, i.e.\ $R(\pa_{\m{x}})$ is the differential operator which one gets by formally replacing all variables $x_j$ in the polynomial $R({\m{x}})$ by $\pa_{x_j}$. Then the following identity holds in the distributional sense
\[R(\pa_{\m{x}})[\del(\m{x})]\, Q(\m{x}) = \left(Q(-\pa_{\m{x}})[R]\right)(\pa_{\m{x}}) \, [\del(\m{x})]. \]
\end{lem}
\pf
It is enough to prove this result in the case where $R$ and $Q$ are polynomials in one variable $x$, i.e.\ $m=1$. Consider $R=x^j$ and $Q=x^k$. Then
\[R(\pa_{x})[\del({x})]\, Q({x})=\del^{(j)}(x) \; x^k =\begin{cases} 0, & k>j,\\ (-1)^k\,\frac{j!}{(j-k)!} \, \del^{(j-k)}(x), & k\leq j.  \end{cases}\]
This coincides with the action on $\del(x)$ of the Fischer dual of the polynomial $(-1)^k \pa_x^k [x^j]=Q(-\pa_x)[R]$, which proves the result. $\hfill\square$

On account of the above lemma, the distribution in (\ref{Dist_inX}) coincides with the action on $\del(\m{x})$ of the Fischer dual of the polynomial
\[S(\m{x}, \m{y}) = \Phi_{m-1}(\|\m{y}\|^2-\langle \pa_{\m{x}},{\m{y}}\rangle^2) [ \Phi_{m}(\|\m{x}\|^2) ]. \] 
If we denote $c_{j,k}=\frac{2\pi^{k/2}}{4^j j! \Gam(j+k/2)}$, we may write
\begin{equation}\label{BigS}
S(\m{x}, \m{y}) = \sum_{\el=0}^\infty c_{\el,m-1} (\|\m{y}\|^2-\langle \pa_{\m{x}},{\m{y}}\rangle^2)^\el \,  \Phi_{m}(\|\m{x}\|^2) =  \sum_{\el=0}^\infty c_{\el,m-1} S_\el
\end{equation}
where 
\begin{align}\label{S_l}
S_\el &= (\|\m{y}\|^2-\langle \pa_{\m{x}},{\m{y}}\rangle^2)^\el \,  \Phi_{m}(\|\m{x}\|^2)=  \sum_{j=0}^\el (-1)^j \binom{\el}{j} \|\m{y}\|^{2(\el-j)} \, \langle \pa_{\m{x}},{\m{y}}\rangle^{2j}  [\Phi_{m}(\|\m{x}\|^2)]\nonumber \\
&=  \sum_{j=0}^\el \sum_{k=0}^\infty  (-1)^j \binom{\el}{j} c_{k+j,m}  \|\m{y}\|^{2(\el-j)} \; \langle \pa_{\m{x}},{\m{y}}\rangle^{2j} \left[\|\m{x}\|^{2k+2j}\right].
\end{align}
%
In order to compute $ \langle \pa_{\m{x}},{\m{y}}\rangle^{2j} \left[\|\m{x}\|^{2k+2j}\right]$ we need the following results.
\begin{lem}\label{Lem8}
Let $B=-\left( \m{x} \wedge \m{y}\right)^2=\|\m{x}\|^2 \|\m{y}\|^2 -\langle {\m{x}}, \m{y} \rangle^2$. Then 
\begin{itemize}
\item[$i)$] $ \langle \pa_{\m{x}}, \m{y} \rangle [B]=0$,
\item[$ii)$] $\langle \pa_{\m{x}}, \m{y} \rangle^2 \left[\|\m{x}\|^{2(k+1)}\right] = 4(k+1) \left[ \left(k+\frac{1}{2}\right) \|\m{y}\|^2 \|\m{x}\|^{2k}-k B\|\m{x}\|^{2(k-1)} \right]$.
\end{itemize}
\end{lem}
\pf
\begin{itemize}
\item[$i)$] It is clear that $\langle \pa_{\m{x}}, \m{y} \rangle [B]= \|\m{y}\|^2\langle \pa_{\m{x}}, \m{y} \rangle \left[\|\m{x}\|^2\right] - \langle \pa_{\m{x}}, \m{y} \rangle \left[\langle {\m{x}}, \m{y} \rangle^2\right]= 2 \|\m{y}\|^2 \langle {\m{x}}, \m{y} \rangle- 2 \|\m{y}\|^2 \langle {\m{x}}, \m{y} \rangle=0$.
 \item[$ii)$] Observe that
 \[\langle \pa_{\m{x}}, \m{y} \rangle \left[\|\m{x}\|^{2(k+1)}\right]=\sum_{j=1}^m y_j \, \pa_{x_j}\left[\|\m{x}\|^{2(k+1)}\right] = 2(k+1)   \sum_{j=1}^m y_j x_j \|\m{x}\|^{2k}= 2(k+1)\langle {\m{x}}, \m{y} \rangle  \|\m{x}\|^{2k}. \] 
 Then,
 \begin{align*}
 \langle \pa_{\m{x}}, \m{y} \rangle^2 \left[\|\m{x}\|^{2(k+1)}\right] &=  2(k+1) \, \langle \pa_{\m{x}}, \m{y} \rangle \left[ \langle {\m{x}}, \m{y} \rangle  \|\m{x}\|^{2k\phantom{)}}\right]\\
& =  2(k+1) \left[ \langle \pa_{\m{x}}, \m{y} \rangle \left[\langle {\m{x}}, \m{y} \rangle\right] \, \|\m{x}\|^{2k} +  \langle {\m{x}}, \m{y} \rangle \,\langle \pa_{\m{x}}, \m{y} \rangle\left[\|\m{x}\|^{2k}\right] \right]\\
&= 2(k+1) \left[  \|\m{y}\|^{2} \|\m{x}\|^{2k} +  2k \langle {\m{x}}, \m{y} \rangle^2 \|\m{x}\|^{2(k-1)} \right].
 \end{align*}
 Substitution of $\langle {\m{x}}, \m{y} \rangle^2=\|\m{x}\|^2 \|\m{y}\|^2-B$ into the previous formula yields the desired result. $\hfill\square$
\end{itemize}
\begin{lem}
The following identity holds for $j,k\in\N\cup\{0\}$  
\begin{equation}\label{DelM}
 \langle \pa_{\m{x}}, \m{y} \rangle^{2j} \left[\|\m{x}\|^{2k+2j}\right] = \frac{4^j(k+j)!}{\Gam\left(k+\frac{1}{2}\right)} \sum_{r=0}^{\min(j,k)} (-1)^r \binom{j}{r} \frac{\Gamma(k+j-r+\frac{1}{2})}{(k-r)!}\|\m{y}\|^{2(j-r)} B^r \|\m{x}\|^{2(k-r)}.
\end{equation}
\end{lem}
\pf 
We proceed by induction over $j\in\N\cup\{0\}$ provided that $k\in\N\cup\{0\}$ is arbitrary. The initial case $j=0$ can be trivially proved to be true. Now assume formula (\ref{DelM}) to be true for some $j\in\N\cup\{0\}$ and any $k\in \N\cup\{0\}$. Let us prove that (\ref{DelM}) also holds for $j+1$.

\noindent It is easily seen that $ \langle \pa_{\m{x}}, \m{y} \rangle^{2j+2} \big[\|\m{x}\|^{2k+2j+2}\big]=  \langle \pa_{\m{x}}, \m{y} \rangle^2  \langle \pa_{\m{x}}, \m{y} \rangle^{2j} \big[\|\m{x}\|^{2(k+1)+2j}\big]$. Using the induction hypothesis  for the pair $(j,k+1)$ and Lemma \ref{Lem8} we have
\begin{multline*}
 \langle \pa_{\m{x}}, \m{y} \rangle^{2j+2} \big[\|\m{x}\|^{2k+2j+2}\big] \\= \frac{4^j(k+1+j)!}{\Gam\left(k+1+\frac{1}{2}\right)} \sum_{r=0}^{\min(j,k+1)} \hspace{-.2cm}(-1)^r \binom{j}{r} \frac{\Gam(k+1+j-r+\frac{1}{2})}{(k+1-r)!} \|\m{y}\|^{2(j-r)} B^r   \langle \pa_{\m{x}}, \m{y} \rangle^2 \big[\|\m{x}\|^{2(k+1-r)}\big].
 \end{multline*}
From now on, we assume $j < k$ and therefore $\min(j,k+1)=j$. The proof in the cases where $j\geq k$ runs along similar lines. We recall from Lemma \ref{Lem8} that
\[ \langle \pa_{\m{x}}, \m{y} \rangle^2 \big[\|\m{x}\|^{2(k+1)-2r}\big]= 4(k-r+1) \left[ \left(k-r+1/2\right) \|\m{y}\|^2 \|\m{x}\|^{2(k-r)}-(k-r) B\|\m{x}\|^{2(k-r-1)} \right].\]
%
Then,
\begin{align*}
&\langle \pa_{\m{x}}, \m{y} \rangle^{2j+2} \big[\|\m{x}\|^{2k+2j+2}\big] \\
&= \frac{4^{j+1}(k+1+j)!}{\Gam\left(k+\frac{3}{2}\right)} \sum_{r=0}^{j} (-1)^r \binom{j}{r} \frac{\Gam(k+j-r+\frac{3}{2})}{(k-r)!} \left(k-r+1/2\right)\|\m{y}\|^{2(j-r+1)} B^r \|\m{x}\|^{2(k-r)}\\
&\phantom{=}+ \frac{4^{j+1}(k+1+j)!}{\Gam\left(k+\frac{3}{2}\right)} \sum_{r=1}^{j+1} (-1)^r \binom{j}{r-1} \frac{\Gam(k+j-r+\frac{5}{2})}{(k-r)!} \|\m{y}\|^{2(j-r+1)} B^r \|\m{x}\|^{2(k-r)}\\
&=\frac{4^{j+1}(k+1+j)!}{\Gam\left(k+\frac{3}{2}\right)} \left( \cit_0 \|\m{y}\|^{2(j+1)}  \|\m{x}\|^{2k} + \sum_{r=1}^{j} (\cit_r+\lan_r)\|\m{y}\|^{2(j-r+1)} B^r \|\m{x}\|^{2(k-r)} + \lan_{j+1}  B^{j+1} \|\m{x}\|^{2(k-j-1)}\right)
\end{align*}
%
where $\cit_{r}=(-1)^r \binom{j}{r} \frac{\Gam(k+j-r+\frac{3}{2})}{(k-r)!} \left(k-r+\frac{1}{2}\right)$ and $\lan_r=(-1)^r \binom{j}{r-1} \frac{\Gam(k+j-r+\frac{5}{2})}{(k-r)!}$. Moreover one has
\begin{align*}
\cit_r+\lan_r &= (-1)^r  j! \, \frac{\Gam(k+j-r+\frac{3}{2})}{(k-r)!} \left(\frac{k-r+\frac{1}{2}}{r! (j-r)!} + \frac{k+j-r+\frac{3}{2}}{(r-1)! (j+1-r)!}\right)\\
&= (-1)^r  j! \, \frac{\Gam(k+j-r+\frac{3}{2})}{(k-r)!}  \frac{(j+1)(k+\frac{1}{2})}{r! (j+1-r)!} \\
&= (-1)^r \left(k+\frac{1}{2}\right)  \binom{j+1}{r} \frac{\Gam(k+j-r+\frac{3}{2})}{(k-r)!} .
\end{align*}
Finally, since $\cit_0=\left(k+\frac{1}{2}\right) \frac{\Gam(k+j+\frac{3}{2})}{k!} $ and $\lan_{j+1}= (-1)^{j+1} \left(k+\frac{1}{2}\right) \frac{\Gam(k+\frac{1}{2})}{(k-j-1)!} $, we conclude
\[\langle \pa_{\m{x}}, \m{y} \rangle^{2j+2} \big[\|\m{x}\|^{2k+2j+2}\big]= \frac{4^{j+1}(k+1+j)!}{\Gam\left(k+\frac{1}{2}\right)} \sum_{r=0}^{j+1} (-1)^r \binom{j+1}{r} \frac{\Gam(k+j-r+\frac{3}{2})}{(k-r)!} \|\m{y}\|^{2(j+1-r)} B^r \|\m{x}\|^{2(k-r)},\]
which completes the proof.
$\hfill\square$

Going back to (\ref{S_l}), we obtain the following results when substituting formula (\ref{DelM}).
\begin{align}\label{S-l2}
S_\el &=  \sum_{j=0}^\el \sum_{k=0}^\infty  (-1)^j \binom{\el}{j} c_{k+j,m}  \|\m{y}\|^{2(\el-j)} \; \langle \pa_{\m{x}},{\m{y}}\rangle^{2j} \left[\|\m{x}\|^{2k+2j}\right] \nonumber\\
&=  \sum_{j=0}^\el \sum_{k=0}^\infty  \sum_{r=0}^{\min(j,k)} (-1)^{j+r} \binom{\el}{j} \binom{j}{r} c_{k+j,m}  \frac{4^j (k+j)! \Gam(k+j-r+\frac{1}{2})}{\Gam\left(k+\frac{1}{2}\right) (k-r)!} \|\m{y}\|^{2(\el-r)} B^r \|\m{x}\|^{2(k-r)} \nonumber\\
&=  \sum_{k=0}^\infty   \sum_{r=0}^{\min(\el,k)} \sum_{j=r}^\el  \frac{2 \pi^{m/2} \el!}{4^k \Gam(k+\frac{1}{2})} \frac{(-1)^{j+r} \, \Gam(k+j-r+\frac{1}{2})}{(\el-j)!\, (j-r)! \, \Gam(k+j+\frac{m}{2})} \; \frac{\|\m{x}\|^{2(k-r)}}{(k-r)!} \, \frac{B^r}{r!} \, \|\m{y}\|^{2(\el-r)} \nonumber \\
&=  \sum_{k=0}^\infty   \sum_{r=0}^{\min(\el,k)} (-1)^r \frac{2 \pi^{m/2} \el!}{4^k \Gam(k+\frac{1}{2})} \, \frac{\|\m{x}\|^{2(k-r)}}{(k-r)!} \, \frac{B^r}{r!} \, \|\m{y}\|^{2(\el-r)}  \left(\sum_{j=r}^\el \frac{(-1)^{j} \, \Gam(k+j-r+\frac{1}{2})}{(\el-j)!\, (j-r)! \, \Gam(k+j+\frac{m}{2})}\right).
\end{align}
%
On the other hand, it can be easily verified that 
\begin{align}\label{HypInterm}
\sum_{j=r}^\el \frac{(-1)^{j} \, \Gam(k+j-r+\frac{1}{2})}{(\el-j)!\, (j-r)! \, \Gam(k+j+\frac{m}{2})} &= \frac{(-1)^r \Gam\left(k+\frac{1}{2}\right) }{(\el-r)! \, \Gam(k+r+\frac{m}{2})} \; {}_2F_1\left(k+\frac{1}{2}, r-\el; k+r+\frac{m}{2};1\right) \nonumber \\
&= \frac{(-1)^r \, \Gam\left(k+\frac{1}{2}\right) \Gam(\el+\frac{m-1}{2})}{(\el-r)! \, \Gam(r+\frac{m-1}{2}) \, \Gam(k+\el+\frac{m}{2})},
\end{align}
where ${}_2F_1(a, b; c; z)=\sum_{j=0}^\infty \frac{(a)_j (b)_j}{(c)_j} z^j$ denotes the hypergeometric function {with} parameters $a,b,c\in \R$ in the variable $z$ and $(q)_j=\begin{cases} 1, & j=0,\\ q(q+1)\cdots (q+j-1), & j>0, \end{cases}$ is the so-called rising Pochhammer symbol, see \cite{MR1688958}. The last equality in (\ref{HypInterm}) is obtained using the Gauss theorem for hypergeometric functions that yields (see \cite{MR1688958})
\[{}_2F_1\left(k+\frac{1}{2}, r-\el; k+r+\frac{m}{2};1\right) = \frac{\Gam(k+r+\frac{m}{2}) \Gam(\el+\frac{m-1}{2})}{\Gam(r+\frac{m-1}{2}) \Gam(k+\el+\frac{m}{2})}.\]
Substituting (\ref{HypInterm}) into (\ref{S-l2}) we obtain
\[
S_\el =  \sum_{k=0}^\infty   \sum_{r=0}^{\min(\el,k)}  \frac{2 \pi^{m/2} \el! \, \Gam(\el+\frac{m-1}{2})}{4^k \Gam(r+\frac{m-1}{2}) \, \Gam(k+\el+\frac{m}{2})} \, \frac{\|\m{x}\|^{2(k-r)}}{(k-r)!} \, \frac{B^r}{r!} \, \frac{\|\m{y}\|^{2(\el-r)}}{(\el-r)!}  .
\]
Substituting this into (\ref{BigS}) yields
\begin{align*}
S(\m{x}, \m{y}) &= \sum_{\el=0}^\infty c_{\el,m-1} S_\el = \sum_{k,\el=0}^\infty   \sum_{r=0}^{\min(\el,k)}  \frac{2 \pi^{\frac{m}{2}} \, 2 \pi^{\frac{m-1}{2}}}{4^{k+\el}  \Gam(k+\el+\frac{m}{2})\, \Gam(r+\frac{m-1}{2})} \, \frac{\|\m{x}\|^{2(k-r)}}{(k-r)!} \, \frac{B^r}{r!} \, \frac{\|\m{y}\|^{2(\el-r)}}{(\el-r)!}.
\end{align*}
The above sum cam be rewritten in terms of the subindex $s=k+\el$ as
\begin{align*}
S(\m{x}, \m{y}) &=  \sum_{s=0}^\infty   \sum_{r=0}^{\lfloor \frac{s}{2} \rfloor} \sum_{k=r}^{s-r} \frac{2 \pi^{\frac{m}{2}} \, 2 \pi^{\frac{m-1}{2}}}{4^{s}  \Gam(s+\frac{m}{2})\, \Gam(r+\frac{m-1}{2})} \, \frac{\|\m{x}\|^{2(k-r)}}{(k-r)!} \, \frac{B^r}{r!} \, \frac{\|\m{y}\|^{2(s-k-r)}}{(s-k-r)!} \\
&= \sum_{s=0}^\infty   \sum_{r=0}^{\lfloor \frac{s}{2} \rfloor}    \frac{2 \pi^{\frac{m}{2}} \, 2 \pi^{\frac{m-1}{2}}}{4^{s}  \Gam(s+\frac{m}{2})\, \Gam(r+\frac{m-1}{2})} \, \frac{B^r}{r!} \left( \sum_{k=r}^{s-r}  \frac{\|\m{x}\|^{2(k-r)}}{(k-r)!}  \, \frac{\|\m{y}\|^{2(s-k-r)}}{(s-k-r)!} \right) \\
&=  \sum_{s=0}^\infty   \sum_{r=0}^{\lfloor \frac{s}{2} \rfloor}    \frac{2 \pi^{\frac{m}{2}} \, 2 \pi^{\frac{m-1}{2}}}{4^{s}  \Gam(s+\frac{m}{2})\, \Gam(r+\frac{m-1}{2})} \, \frac{A^{s-2r}}{(s-2r)!} \,  \frac{B^r}{r!}, 
\end{align*}
where $A=\|\m{x}\|^2+\|\m{y}\|^2$ and $B=\|\m{x}\|^2 \|\m{y}\|^2 -\langle {\m{x}}, \m{y} \rangle^2$.

Finally, using our first relation (\ref{PrimFormS2}) and Lemma \ref{FiscDual}, we rewrite the Pizzetti formula  (\ref{PizStie2_1}) as follows.
%
\begin{teo}\label{PizT}{\bf [Pizzetti formula for $\textup{St}(m,2)$]}
For any polynomial $P(\m{x},\m{y}): \R^m\times\R^m\fd \C$ it holds that 
\begin{align*}
\int_{\textup{St}(m,2)}P(\m{x}, \m{y})  \, dS_{\m{x},\m{y}} &= S(\pa_{\m{x}}, \pa_{\m{y}}) [P] \big|_{\m{x}=0 \atop \m{y}=0}\\
&= A_{m} A_{m-1} \sum_{s=0}^\infty   \sum_{r=0}^{\lfloor \frac{s}{2} \rfloor} \frac{\Gam\left(\frac{m}{2}\right)}{4^{s}  \Gam(s+\frac{m}{2}) } \frac{\Gam\left(\frac{m-1}{2}\right)}{\Gam(r+\frac{m-1}{2})}  \frac{{\bf A}^{s-2r}}{(s-2r)!} \frac{{\bf B}^{r}}{r!} [P]\Bigg|_{\m{x}=0 \atop \m{y}=0}
\end{align*}
with ${\bf A}=\Del_{\m{x}} + \Del_{\m{y}}$, ${\bf B}= \Del_{\m{x}}\Del_{\m{y}}-\langle \pa_{\m{x}}, \pa_{\m{y}}\rangle^2$ and $A_j=\frac{2\pi^{j/2}}{\Gam(j/2)}$ the surface area of the $(j-1)$ dimensional unit sphere $\Sa^{j-1}$.
\end{teo} 


\bibliographystyle{abbrv}

\end{document}